# Ferroelectric Polycrystals: Structural and microstructural levers for property-engineering via domain-wall dynamics

J. Schultheiß[1,2*], G. Picht[3], J. Wang[4], Y. A. Genenko[5], L.Q. Chen[4], J.E. Daniels[6], and J. Koruza[7*]

[1] Department of Materials Science and Engineering, Norwegian University of Science and Technology (NTNU), 7034 Trondheim, Norway

[2] Experimental Physics V, University of Augsburg, 86159 Augsburg, Germany

[3] Corporate Sector Advanced Technology, Advanced Technologies for Chemical and Biological Systems, Robert Bosch GmbH, 71272 Renningen, Germany

[4] Department of Materials Science and Engineering, The Pennsylvania State University, University Park, Pennsylvania 16802, USA

[5] Institute of Materials Science, Technische Universität Darmstadt, Otto-Berndt-Straße 3, 64287, Darmstadt, Germany

[6] School of Materials Science and Engineering, University of New South Wales, 2052 Sydney, Australia

[7] Institute for Chemistry and Technology of Materials, Graz University of Technology, 8010 Graz, Austria
*corresponding authors: jan.schultheiss@ntnu.no, jurij.koruza@tugraz.at

**Abstract**

Ferroelectrics have a spontaneous electrical polarization that is arranged into domains and can be reversed by an externally applied field. This high versatility makes them useful in enabling components such as capacitors, sensors, and actuators. The key to tuning their dielectric, piezoelectric, and electromechanical performance is to control the domain structure and the dynamics of the domain walls. In fixed compositions, this is often realized by chemical doping. In addition, structural and microstructural parameters, such as grain size, degree of crystallographic texture or porosity play a key role. A major breakthrough in the field came with the fundamental understanding of the link between the local electric and mechanical driving forces and domain wall motion. Here, the impact of structure and microstructure on these driving forces is reviewed and an engineering toolbox is introduced. An overview of advances in the understanding of domain wall motion on the micro- and nanoscale is provided and discussed in terms of the macroscopic functional performance of polycrystalline ferroelectrics/ferroelastics. In addition, a link to theoretical and computational models is established. The review concludes with a discussion about beyond state-of-the-art characterization techniques, new approaches, and future directions toward non-conventionally ordered ferroelectrics for next-generation nanoelectronics and energy-storage applications.

**Content**

# 1. Introduction

Ever since their discovery over 100 years ago, [1] ferroelectric materials have been a fascinating research topic. Their numerous functional and physical properties make them indispensable in various technological products, such as actuators, capacitors, sensors, and transducers. [2, 3] The field is continuously growing due to ongoing innovations, ranging from nanotechnology [4] to energy storage [5] and medical applications [6]. The intriguing physical properties of these materials are related to a spontaneous polarization, which forms below the Curie temperature on the unit cell level and can be reversed by an external field. A coherent alignment of unit cells with uniform polarization direction is referred to as a ferroelectric domain. [7] Domains naturally form in ferroelectric materials to minimize the electrostatic and elastic energy. Ferroelectric 180° domain walls separate volumes where the spontaneous polarization vectors are parallel but of opposite sign, while ferroelectric/ferrolastic non-180° domain walls separate regions with non-parallel polarization vectors (further details on domain types are provided in section 4.1.1.1).

Domain walls and their dynamics are heavily researched, since they co-determine many of the functional properties of ferroelectric materials, such as the switchable polarization, dielectric, or piezoelectric response. In this frame, the functional properties can be disentangled into intrinsic and extrinsic contributions. [7] The intrinsic response is related to the unit cell, i.e., distortions of the crystal lattice in response to some perturbation. The extrinsic response is generally associated with electric and/or mechanical field-induced displacement of domain walls or phase transformations. [8] Extrinsic ferroelectric/ferroelastic domain wall movement may contribute up to 80% to the dielectric and piezoelectric properties in polycrystals [9] and the functional properties of ferroelectrics/ferroelastics can be tuned via engineering the domain structure and the response of ferroelectric/ferroelastic domain walls to an applied external stimulus. In comparison to single crystalline counterparts, intergranular interactions, i.e., elastic strains originating from a rigid coupling between individual grains, and grain boundaries play a key role in polycrystals, impacting intrinsic and extrinsic contributions. [10] The dynamics of domain walls depend on the chemical composition of the polycrystals. Material-specific review articles exist for $BaTiO_3$ (BT)-based, [11] $BiFeO_3$ (BFO)-based, [12] and $(K,Na)NbO_3$ (KNN)-based [13] materials. Beyond the initial parent composition, chemical doping is often used to control domain wall dynamics. By creating defect complexes that interact with the small- and large-scale movement of domain walls, properties can be varied. [14-17]

Here, we move beyond the chemical doping strategy and discuss the general influence of structural and microstructural parameters on the functional properties of polycrystalline ferroelectric/ferroelastic ceramics with a focus on the dynamics of domain walls. First, the main driving forces for domain wall movement in single crystalline materials will be explained (section 2.1), followed by describing the impact of polycrystallinity on these forces (section 2.2). Next, selected imaging techniques will be summarized, driving the field forwards due to their increasingly accurate capability of direct mapping and quantifying the static domain structure of ferroelectrics and the impact of an external stimulation on domain wall motion (section 3). In the main section, the impact of structural and microstructural parameters, including crystal structure (section 4.1), grain size, and orientation (section 4.2), as well as porosity (section 4.3) on domain wall dynamics and functional properties will be reviewed with a focus on BT and $Pb(Zr,Ti)O_3$ (PZT) based model systems. Selected computational methods which allow to shed light on the interplay between the microstructure and domain wall dynamics on different length scales will be discussed (section 5). We end our review with an outlook on advanced characterization techniques and outline their potential to disentangle the complex interplay between structure and domain wall dynamics (section 6.1), more recent approaches to tailor domain wall dynamics using extended defects (section 6.2), and polycrystalline systems with non-classical ferroelectric order (section 6.3).

# 2. Driving forces for domain wall dynamics in ferroelectrics/ferroelastics

In our review the impact of structure and microstructure on enhancing and inhibiting domain wall movement will be discussed and linked to functional properties of ferroelectric/ferroelastic materials. To do this, we will first address the question of how domain walls move in the single crystalline model case, and next explain the impact of the more complex boundary conditions found in polycrystalline materials. While the focus of our review is on the mechanisms, the understanding of domain wall dynamics in polycrystalline ferroelectrics/ferroelastics was driven forward by advances in characterization techniques. For a more detail insight and application examples of bulk-averaging characterization techniques, we refer the reader to refs. [8] and [10]. Due to their capability to cover a broad electric field and time range, we focus on dynamic, i.e., time-dependent measurements, which enable distinction between individual types of domain wall movement. In addition to this, an important insight on the interplay between structure/microstructure and domain wall dynamics will be gained from frequency, [18] temperature, [19] and both above- and sub-coercive field-dependent measurements [20, 21]. Furthermore, X-Ray (XRD) [10, 22-30] and neutron [31-33] diffraction provide in-situ capabilities and access to the domain switching fraction and elastic strains averaged over different grain families, which is one of the keys towards disentangling the intergranular interactions in polycrystalline ferroelectrics/ferroelastics.

## 2.1. Single crystals

To reveal the main driving forces for domain wall movement, we first recall the polarization reversal process in a single crystal, schematically illustrated in Figure 1a. Application of an electric field to a monodomain crystal results in the nucleation of new domains, their growth through the crystal, followed by sidewise expansion and domain coalescence. A more detailed description of the individual steps is provided in ref. [34], while a review of the detailed mechanisms for domain wall movement is given in ref. [15]. In addition to this established mechanism, polarization reversal may occur without domains, if the domain nucleation can be suppressed. This requires ferroelectric single crystals of high purity and electric fields far above the coercive field need to be applied under a high ramp rate. This exotic scenario was experimentally observed in ref. [35] and a more detailed explanation is provided in ref. [36].

To understand the main driving force for domain wall movement, we display electric-field-dependent velocity studies of 180° domain walls in BT single crystals. [37-39] Pioneering experimental results [40-46] are summarized in Figure 1b. A strong dependence of the velocity on the applied electric field amplitude can be observed. Two regimes are apparent: an exponential relationship was found for low electric fields ($E$<0.1 kV/mm), [40] while a power law was proposed for high electric fields ($E$>0.1 kV/mm). [44] The Merz law describes the exponential relationship between the domain wall velocity, $v(E)$, and the applied electric field, $E$ [37]:

$$v(E) = v_0 \cdot exp[-(E_a/E)] \qquad \text{Equation 1}$$

The parameter $E_a$ is termed the activation field, which is linked to the material-specific activation barrier, $U$, via

$$E_a = E_{c_0} \cdot \frac{U}{k_b \cdot T}. \qquad \text{Equation 2}$$

Here, $E_{c_0}$ is a critical field at which domain wall depinning occurs at 0 K, $k_b$ is the Boltzmann constant, and *T* is the temperature. A ferroelectric domain wall is moving if the interplay between the thermal fluctuation and the external fields is sufficient to overcome the activation barrier. Mechanistically, the domain wall does not move as a whole but by small jumps, [42, 47-51] the so called Barkhausen jumps [52]. In an ideal defect-free single crystal, the height of the potential barrier, and thus the parameter $E_a$, is defined by the Peierls potential of the lattice. [15, 53-55] The local energy landscape depends on chemical defects, including electronic charge carriers, ionic defects, defect dipoles, and defect complexes. The interplay between chemical defects and domain wall dynamics has been studied experimentally [56, 57], as well as theoretically [58, 59]. Since the impact of chemical defects on domain wall dynamics goes beyond the scope of this review, we refer the reader to refs. [14-17] for a more complete literature overview.

While experimental observations on single crystalline materials with a monodomain starting state confirmed the idealized scenario displayed in Figure 1a, [34, 60, 61] domain wall dynamics become more complicated if the starting state deviates from this scenario. For example, polarization reversal occurs via successive non-180° domain switching steps, if non-180° domain walls are present before application of the electric field. [62, 63] The movement of non-180° domain walls in single crystals was confirmed by electromechanical, [64, 65], acoustical, [66, 67] optical, [68-70] and diffraction [71, 72] studies. These works outline that besides the electric field, as displayed in Figure 1 and Equation 1, mechanical driving forces also play a significant role for the movement of domain walls, related to their additional ferroelastic nature. Mechanical stress is known to further modify the height of the potential barrier and thus impact the parameter $E_a$, [73, 74] acting as an additional driving or hindering force for domain wall movement. [75, 76] Furthermore, domain wall movement was found to be impacted by other obstacles, such as other domain walls [77, 78] or electrode defects [79, 80].

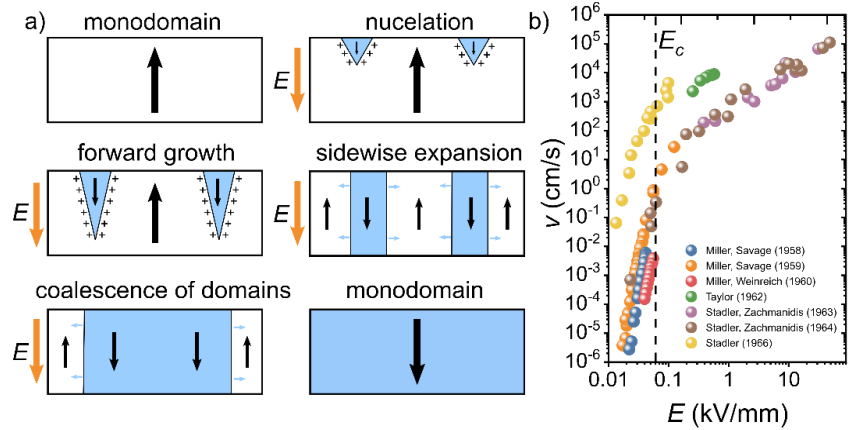

*Figure 1:* Polarization reversal in a monodomain single crystal. a) Schematic representation of the polarization reversal process under an applied electric field, E. The mechanism includes the nucleation of new domains, forward growth, sidewise expansion, and domain coalescence. [34] b) Field-dependent velocity of 180° domain walls, v, in single crystalline BT during the sidewise expansion. [40-46]. The coercive field of $E_c$=0.06 kV/mm of a BT single crystal, obtained from polarization hysteresis measurements, is indicated for comparison by a vertical dashed line. [81]

## 2.2. Polycrystals

Similar to their single crystalline counterparts, 180° [82-84] and non-180° [85-88] domain wall movements were found in polycrystalline ceramics. A breakthrough in understanding the driving forces for domain wall movement in polycrystalline materials was the revelation of the sequence of domain switching events from time-dependent dynamic measurements. This was accomplished using a synergetic approach combining high-energy diffraction, dynamic polarization, and strain measurements, [89, 90] as well as stochastic modelling [91] on a poled polycrystalline ferroelectric/ferroelastic material. Ferroelectric 180° and ferroelectric/ferroelastic non-180° domain switching events can be disentangled based on their individual contribution to the time evolution of polarization and strain under a constant electric field pulse. [91] A typical time-dependent response of the switched polarization, Δ*P*, and strain, *S*, of the material to a high-voltage pulse applied anti-parallel to the poling direction is schematically displayed in Figure 2a. Based on this, a sequential mechanistic description of domain dynamics featuring three different regimes has been revealed (Figure 2b): 1) non-180° domain wall movement, 2) main switching phase, and 3) creep-like domain wall movement. First, we will discuss the three regimes mechanistically (section 2.2.1), before we will consider the impact of the local electric and mechanical driving forces on the dynamics of domain walls (section 2.2.2) in detail.

### 2.2.1. Sequence of domain switching events

As a consequence of high lattice strain after electric poling and a four times lower domain wall energy of non-180° compared to 180° walls, [92] a high density of non-180° domain walls exists in the poled state of a polycrystalline material (regime 0 in Figure 2). [93-97] Application of an external electric field antiparallel to the poling direction results in a rapid shrinkage of the sample and the development of a negative macroscopic strain due to non-180° domain wall movement (Figure 2a). [10] The driving force for non-180° domain wall movement is the local electric field [98] supported by mechanical driving forces [90] originating from grain-to-grain coupling. [99] Electric and mechanical driving forces are displayed by orange and green arrows in Figure 2b, respectively. At the end of regime 1 a critical state within the grains is achieved, defined by the interplay between local mechanical and electric fields. This triggers the main switching phase (regime 2 in Figure 2), [89] where more than 60% of Δ*P* reverses. [89, 100, 101] Mechanistically, two different scenarios, as highlighted in i) and ii) in Figure 2b, may occur: i) 180° domain walls nucleate and move sidewise similar to the situation in single crystals (Figure 1a) [102], ii) 180° domain switching occurs by synchronized non-180° domain wall dynamics. The latter requires the nucleation of a non-180° domain wall, which moves through the domain structure of a grain. [103, 104] While current experimental techniques do not allow to distinguish between both suggested paths, microstructural arguments favor synchronized non-180° domain wall movement. For example, a reduced Peierls barrier, [49, 92, 105, 106] crystallographic arguments, [107, 108] and reduced activation barriers [109] indicate that non-180° domain wall movement is energetically favorable compared to 180° domain wall movement. At the end of the main switching phase, most domains have reversed, and the material enters regime 3 (3 in Figure 2). Here, the electric field is parallel to the polarization vector and the time-dependent polarization [110] and strain [111] curves (Figure 2a) resemble electric creep. The slow increase of polarization and strain in this regime was attributed to the progressive movement of the ferroelectric/ferroelastic domain walls. [112, 113]

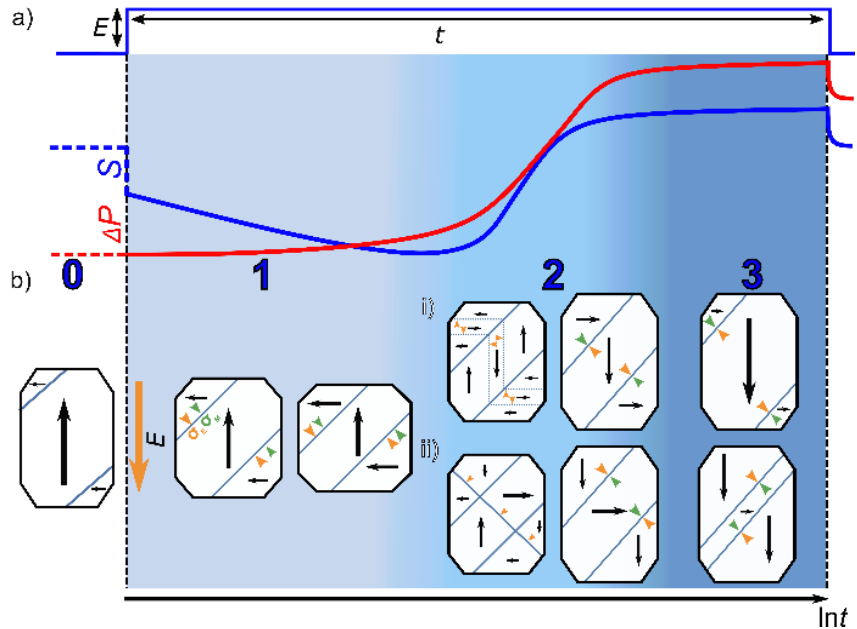

*Figure 2: Domain wall dynamics in ferroelectric/ferroelastic polycrystalline ceramics. a) Schematic response to a high voltage (HV) pulse with height E and time t. Based on the evolution of strain, S, and switched polarization, ΔP, three regimes can be distinguished. The schematic domain response is displayed in b) for a grain with a tetragonal crystal structure. The 90° domain walls are displayed as solid lines, while the 180° domain walls are dashed. 1) Initial non-180° domain wall movement, 2) main-switching phase by a) nucleation and sidewise growth of 180° domains, [102] and b) 180° polarization reversal by a nucleated non-180° domain wall through synchronized non-180° domain-wall movements, [103, 104] and 3) creep-like domain wall movement. Electric and mechanical driving forces are displayed by orange and green arrows, respectively. Reprinted from [89], with permission from Elsevier.*

2.2.2. Local electric and mechanical driving forces for domain wall dynamics

Similar to single crystalline materials (Figure 1) discussed in section 2.1, the local electric field is the main driving force for the polarization reversal process in polycrystalline ferroelectric/ferroelastic ceramics (orange arrow in Figure 2b). In contrast to single crystals, where the value of the local electric field is similar to the externally applied one, [114, 115] the local electric field is inhomogeneously distributed in polycrystals, [83, 116-120] which is relevant to discuss the impact of structure and microstructure on switching dynamics. The local electric field thereby is a projection of the external field onto the local spontaneous polarization vector of each individual domain inside each grain. [119, 120] In a first approach, this distribution can be calculated by considering the dielectric permittivity tensor and the distribution of polarization vectors. [121, 122] The statistical distribution of the local electric field values in polycrystals is roughly symmetrically centered around the value of the external applied field. For the model case of a dense tetragonal $Pb(Zr_{0.515}Ti_{0.485})O_3$ material, [121] the local electric field is enhanced/reduced by ~20% in ~85% of the grains, while strong field deviations exceeding 60% enhancement/reduction occur in <5% of the grains. Crystal structure, [121, 123-125] degree of crystallographic texture [126], stress concentrations near grain boundaries, [127] and porosity [101, 128] were suggested as engineering levers for tuning the local electric field distribution, as will be discussed in more details in chapter 4.

In comparison to single crystalline materials, where the local deformation related to the piezoelectric effect and the movement of ferroelectric/ferroelastic domain walls is unconstrained, [129] in polycrystalline materials the grains are mechanically coupled over a rigid grain boundary, as demonstrated experimentally. [15, 89, 99, 130-136]. Mechanical stress is a direct consequence of the coupling between the grains (green arrow in Figure 2b). [89] The anisotropy of both intrinsic and extrinsic piezoelectric strains, as well as the elastic anisotropy of the system, will cause some grain orientations to drive the response, while others restrict the response due to their requirement for elastic accommodation. A rigorous treatment of these intergranular interactions using the Eshelby approach [137] is provided in ref. [99]. Considerable intergranular stress in the order of 40-100 MPa were reported for polycrystalline ferroelectric/ferroelastics. [138]

## 3. Imaging techniques

Ferroelectric/ferroelastic polycrystals offer various engineering degrees of freedom, including grain boundaries, [139-141] a locally distributed electric field, [83, 120, 128] and 3D confinement effects. [89, 99] One of the key challenges in the field therefore is to develop suitable methods enabling to visualize the static domain structure, i.e., the shape and size of the ferroelectric domains, and to image the impact of external driving forces on domain wall dynamics. Due to their capability to directly quantify and map the correlation between the micro- and domain structure as well as the dynamics of domain walls, we focus on selected imaging techniques in this section, which, in our opinion, are advancing the fundamental understanding of the interplay between structure and microstructure on domain wall dynamics in polycrystalline ferroelectrics/ferroelastics. To further consider the different chemical and mechanical boundary conditions [142-145] that impact structure-property relationships in polycrystalline ferroelectrics, we have structured the section into surface-sensitive (section 3.1) and bulk-sensitive (section 3.2) techniques.

3.1. Surface-sensitive imaging techniques

Classically, domains in ferroelectric/ferroelastic polycrystals were visualized via optical microscopy (Figure 3a). These methods were limited to domain structures with sizes above approximately 0.5 µm. [146] A major achievement in the field was the development of domain imaging techniques pushing the resolution towards the nanoscale. Scanning electron microscopy (SEM) and piezoresponse force microscopy (PFM) have facilitated imaging of the complex coupling between microstructure and domain morphology at length scales well below 100 nm. Selected domain images obtained by SEM and PFM are displayed in Figure 3b and Figure 3c, respectively.

SEM provides a direct overview of the domain structure of several grains. The domains can be revealed either by chemical or thermal etching or by electron channeling contrast imaging of well-polished samples. A detailed overview of the characterization and experimental parameter ranges of ferroelectric domains by SEM is provided in refs. [147, 148]. In comparison to SEM, PFM enables to map the electromechanical properties with nanoscale resolution, facilitating visualization of the ferroelectric domain structure. In addition to mapping the static properties of ferroelectric domains, spectroscopic approaches with nanoscale resolution broaden the measurement capabilities, e.g., by quantifying the impact of electric driving forces on domain wall dynamics. [149] Comprehensive reviews on the working principle of PFM and exemplarily domain structures of ferroelectric materials can be found in refs. [150-153]. Tomographic PFM further extends the analysis into subsurface regions of ferroelectric materials. [154] To obtain even higher resolution, transmission electron microscopy (TEM) directly visualizes atomic positions, facilitating atomic-scale structural resolution of a ferroelectric domain wall (Figure 3d). In addition, photoemission electron microscopy (PEEM), [155] low-energy electron microscopy (LEEM), [156] and confocal Raman spectroscopy, [157] have been explored to widen the experimental parameter range for mapping ferroelectric domain structures.

3.2. Bulk sensitive imaging techniques

Bulk-sensitive microscopy imaging provides a new dimension to the techniques introduced in section 3.1, since static and dynamic information of the domain structure can be readily obtained under natural mechanical, electrostatic, and chemical boundary conditions in the bulk of ferroelectric/ferroelastic polycrystals. In addition, they are a powerful tool to unravel the structure-property relationship since they combine domain imaging with local maps of elastic strain, giving direct access to elasto-morphological correlations in the bulk as a function of an external stimulation.

X-Ray projection ptychography maps the ferroelectric domain structure in thin films, with a spatial resolution down to 5.7 nm. [158] A more detailed explanation of the technique is provided in ref. [159]. Scanning highly coherent beams enables to capture the dynamics of ferroelastic domains in nanowires. [160] For ferroelectric polycrystals, 3D grain mapping techniques, such as 3D X-ray diffraction (3D-XRD) [161, 162] and diffraction contrast tomography, [163, 164] allow a rapid overview of the grain morphology and orientation, and visualize the response of individual grains with a resolution down to 1 µm. Combined with an electric field, this facilitates quantification of the strain response of each individual grain (Figure 3d). [165, 166] Related methods are also developing using neutrons, opening the possibility of probing much larger bulk polycrystalline materials, though the resolution is limited in comparison to XRD-based methods. [167] Dark-field X-Ray microscopy (DFXM) [168-170] allows observation of strain fields (Figure 3f) together with crystallographic orientation maps, [170] with resolutions in the sub-100 nm regime (Figure 3e), [171] slowly approaching the spatial resolution of SPM or SEM methods. For a more detailed insight into DFXM, we refer the reader to refs. [172, 173].

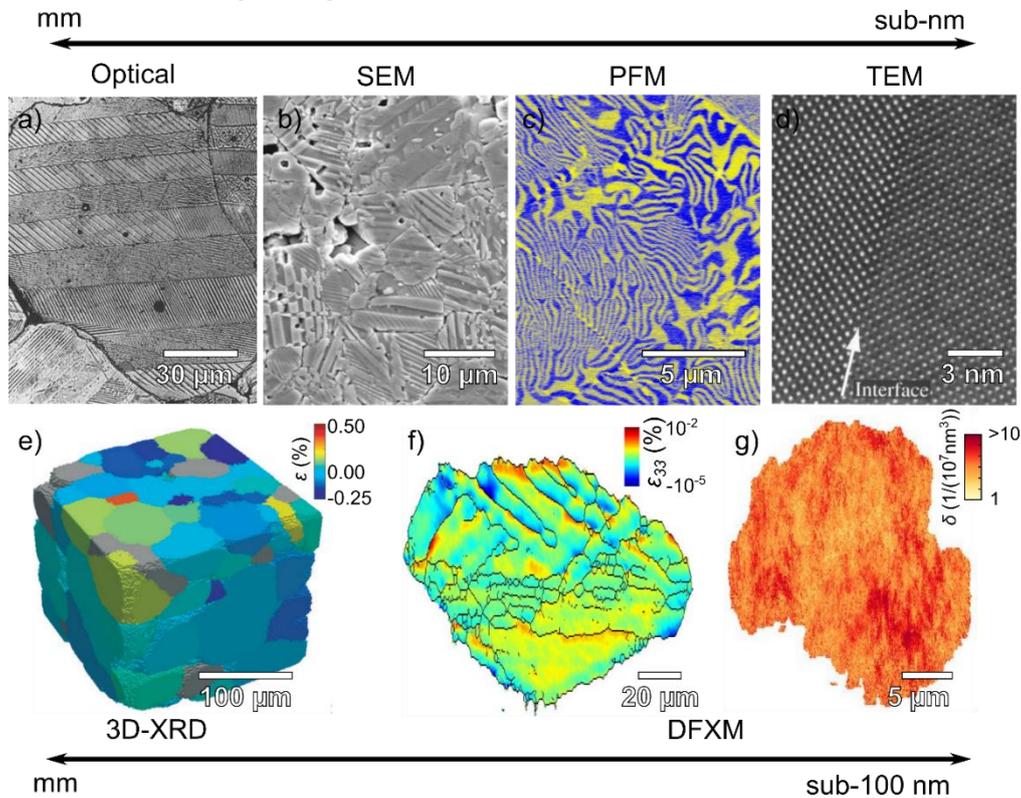

*Figure 3:* Selected techniques which enable multilength scale mapping of ferroelectric domains in polycrystalline ferroelectric/ferroelastic materials, providing information on the domain structure and domain wall dynamics. a) Micrograph of polycrystalline BT obtained by optical microscopy. Reprinted by permission from Springer Nature Customer Service Centre GmbH: [174], Copyright 1990. b) SEM micrograph of an etched surface of $(K,Na)_{0.94}Li_{0.06}NbO_3$ featuring a complex domain structure. Reprinted by permission from Springer Nature Customer Service Centre GmbH: [175], Copyright 2018. c) PFM of polycrystalline $ErMnO_3$ displaying the complex domain structure which forms by the interaction of topologically protected vortex/anti-vortex pairs and elastic strain fields. Reprinted from [176] with permission from Wiley. d) High resolution TEM micrograph of a 90° domain wall in $PbTiO_3$ (PT). [177], by permission of The Japanese Society of Microscopy. e) Grain maps of a polycrystalline BT sample in the virgin state obtained using 3D-XRD. The average strains, $\varepsilon$, of the individual grains are displayed. Reproduced with permission from [165], Copyright 2017, The American Ceramic Society. f) Cross sectional cut through a BT grain obtained by DFXM. Strain concentrations in the vicinity of interfaces, e.g., domain walls (black lines) are displayed. [170] g) Density of domain variants in a slice of a grain of a polycrystalline BCZT material. The number of domains inside a localized diffraction volume of $10^7$ nm³ is displayed. [171]

4. **Structural and microstructural toolbox**

An overview of the degrees of freedom offered by the structure and microstructure is provided in Figure 4. In this review, we disentangle the high interdependency between each of the parameters and clarify the individual impacts on the static domain structure and domain wall dynamics. Note that Figure 4 focuses on structural and microstructural elements, while composition specific approaches, such as doping or defect chemistry, which are frequently applied to control domain dynamics are not included (the interested reader is referred to refs. [14-17]). Section 4.1 deals with intrinsic structural parameters, i.e., the crystal structure and the distortion of the lattice (Figure 4a). Sections 4.2 and 4.3 discuss the impact of the grain boundaries, the grain-orientation relationship, and the porosity on domain dynamics (Figure 4b). The potential of new approaches using extended defects (Figure 4c), including dislocations, secondary phases, and electrical inhomogeneities will be briefly outlined in section 6.2.

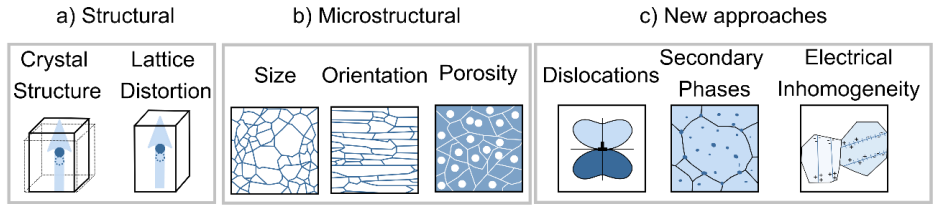

*Figure 4: Structural and microstructural toolbox to tune functional properties in polycrystalline ferroelectric/ferroelastic materials via domain wall dynamics: a) intrinsic structural, b) microstructure-related parameters, and c) new approaches using extended defects.*

### 4.1. Intrinsic structural parameters

While a coexistence of crystallographic phases, e.g., the concept of the morphotropic phase boundary (MPB), is important in terms of an enhancement of dielectric and piezoelectric properties, the exact mechanisms are still under discussion. In this section, we therefore focus on the impact of the individual crystal structures with the goal to establish a mechanistical understanding linking the amount of polarization directions and spontaneous strain to the dynamics of domain walls (section 4.1.1). Next, we review literature on the impact of the magnitude of the lattice distortion away from the parent paraelectric state and evaluate the domain-wall related response (section 4.1.1.2). Finally, the mechanisms are demonstrated via two selected case studies (section 4.1.2), establishing a link between domain wall dynamics and functional performance.

#### 4.1.1. Mechanistic description

##### 4.1.1.1. Tetragonal, orthorhombic, and rhombohedral phases

The polar axes and the amount of equivalent polarization directions vary for the different crystal symmetries as summarized in Table 1. The number of equivalent polarization directions is 6, 12, and 8 for tetragonal, orthorhombic, and rhombohedral structures, respectively. [178] Systems with a coexistence of rhombohedral and tetragonal phases have been described by 14 equivalent polarization directions. [179] The crystal structure impacts the local electric field, which is one of the driving forces for domain wall movement (section 2.2.2). The correlation between crystal structure and electric-field distribution, as quantified by its Full-Width-Half-Maximum (FWHM in Table 1), was measured [124] and calculated [100, 121]. Exemplary calculations found that the distribution is broadest for a tetragonal material (FWHM=0.32), while it is narrowest for a rhombohedral counterpart (FWHM=0.19). Further, a one-to-one correlation between the crystal structure and the switchable polarization exists. In a single crystalline ferroelectric, the entire polarization can be aligned if the external field is applied in the direction of a polar axis, e.g., along [100] in tetragonal materials ($P_{max}/P_S$=1). In polycrystalline materials, the maximal possible fraction $P_{max}/P_S$ is reduced and scales with the amount of the available polarization directions. [180] As displayed in Table 1, values of $P_{max}/P_S$ = 0.866 and $P_{max}/P_S$ = 0.831 were calculated for rhombohedral and tetragonal polycrystalline materials, respectively, while a higher value of $P_{max}/P_S$ = 0.912 was reported for an orthorhombic structure. [119, 121, 181, 182] For comparison, a value of $P_{max}/P_S$ = 0.922 was reported for a tetragonal and rhombohedral phase coexistence. [183, 184] Going beyond the impact of the crystal structure on electric driving forces and switchable polarization, also mechanical driving forces are impacted via the spontaneous strain of the unit cell. In a polycrystalline ferroelectric/ferroelastic material, the electric-field induced movement of non-180° domain walls comes along with the generation of internal stress, [99] which were found to be higher in tetragonal compared to rhombohedral structures. [185, 186]

*Table 1: Impact of crystal structure on domain wall dynamics: Number of equivalent directions, allowed types of non-180° domain walls, maximum possible fraction of single crystal polarization value, which can be achieved in a random polycrystalline state, $P_{max}/P_S$, and FWHM of the local electric field distribution. [After ref. [178]]*

| Symmetry | Polar axis | Number of equivalent pol. directions | Types of non-180° domain walls | $P_{max}/P_s$ (ref. [181, 183]) | FWHM of the local electric field distribution (ref. [121]) |
|---|---|---|---|---|---|
| **Tetragonal** | <001> | 6 | 90° | 0.831 | 0.32 |
| **Orthorhombic** | <110> | 12 | 90° / 60° / 120° | 0.912 | 0.29 |
| **Rhombohedral** | <111> | 8 | 71° / 109° | 0.866 | 0.19 |

##### 4.1.1.2. Lattice distortion away from the parent cubic phase

The magnitude of the distortion of the lattice away from the parent cubic phase mainly impacts local mechanical driving forces (section 2.2.2) as can be best understood if the responses of a single- and a polycrystalline material with a high lattice distortion are compared. A good example is PbTiO$_3$ (PT), which has a high tetragonal lattice distortion of $c/a \sim 1.063$. Macroscopic electric-field-dependent polarization measurements on polycrystalline PT [187, 188] did not reveal the development of a polarization hysteresis loop when an electric or mechanical field was applied, indicating that domain wall movement in polycrystalline PT is suppressed due to mechanical stress originating from grain-to-grain interactions. [189] This is manifested by an unchanged tetragonal domain switching fraction for electric fields as high as 5 kV/mm or under compressive stress as high as 300 MPa. [187] In comparison, measurements on PT single crystals found a low coercive field of 0.64 kV/mm and a high remanent polarization of 74 µC/cm², [190, 191] which is close to the theoretical value. [192]

Mechanistically, mechanical stress in the vicinity of domain walls scales with the lattice distortion. The reason is, that the angle between two adjacent domains deviates from the ideal strain-free 90° configuration of a tetragonal crystal structure [193-195], experimentally resolved for example for PT thin films [196]. For tetragonal materials, the deviation angle $\xi$ thereby depends on the tetragonal $c/a$ lattice distortion and can be calculated as

$$90° - \zeta = \arctan((c/a)^{-1}).  \qquad \text{Equation 3}$$

Calculations for different lattice distortions are in good agreement with experimental results. For example, BT with a low lattice distortion of $c/a$=1.01 exhibits a deviation angle of $\zeta = 0.5°$, [197-199] while PT with a lattice distortion of $c/a$=1.063 has a higher deviation angle of $\zeta = 3.6°$. [200, 201] The origin of the internal compatibility stress at a domain wall is derived in Figure 5a-c.

[202] A drawing of the stress-free domain structure for a tetragonal material is schematically depicted in Figure 5a. This schematic domain structure can be separated into three pieces, as displayed in Figure 5b. Mechanical stress arises when the domain structure is forced back to be exactly 90° (Figure 5c). Large tensile and compressive elastic stress is stored in such junctions between different lamellar regions, as indicated by the arrows in Figure 5c. [203-206] The stress level is plotted as a function of the $c/a$ ratio in Figure 5d for a tetragonal PZT model system and significant stress levels in the GPa regime were calculated for the junction of 90° domains. [207, 208] The blue data points in Figure 5d represent the theoretically calculated deviation angle according to Equation 3, while the orange data points represent experimentally obtained values. Results obtained by DFXM further demonstrate that the elastic stress fields in the vicinity of domain walls are long ranging and can extend up to several micrometer into the adjacent domains (Figure 3b). [170]

Besides at domain walls, stress concentrations are also found near grain boundaries. These arise in all polycrystalline materials, e.g., due to the anisotropy of the thermal expansion coefficient, the random orientation of the grains, [171, 209, 210] and the formation of a spontaneous strain at the phase transition temperature. [129] For ferroelectric materials with a high lattice distortion, e.g., PT or $YMnO_3$, high stress result in microcracks [211, 212] or even the fracture of the polycrystalline material during cooling [213]. The correlation between lattice distortion and stress was systematically studied for La-doped PT. [214] Internal mechanical stress was found to scale with the $c/a$ ratio, as displayed in Figure 5e. While average internal mechanical stress is absent for a lattice distortion close to cubic ($c/a$~1.00), values of approximately 150 MPa were found for a lattice distortion of 1.03. Furthermore, a correlation between the $c/a$ ratio and the static domain structure was observed. TEM micrographs obtained for two representative lattice distortions of 1.007 and 1.030 of a La-doped PT ceramic are displayed in Figure 5e. [215] While strain-induced domains were found to be absent for a $c/a$ ratio close to the cubic parent phase ($c/a$=1.007), a high density of domain walls was observed if the $c/a$ ratio was increased to 1.030. A similar correlation between the lattice distortion and the static domain structure for other systems is also reported in refs. [216, 217].

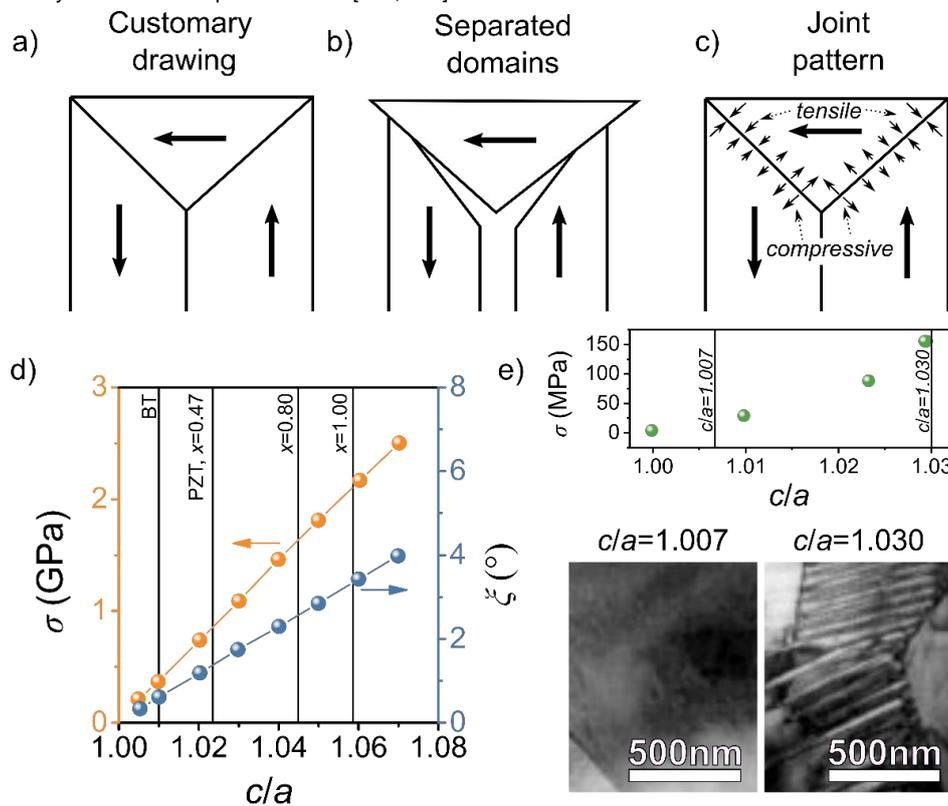

*Figure 5:* Interplay between lattice distortion and microstructure. A possible origin of mechanical stress in the vicinity of the domain walls is exemplarily shown for a simple 2D banded domain structure in a)—c). a) The customary drawing of the domain structure displays the stress-free case. The lattice distortion results in a space mismatch between the domains, as displayed in b). After joining the domain patters, tensile and compressive stress develop at the domain wall, as schematically displayed by the arrows in c). [202] d) The influence of lattice distortion on the maximum stress, $\sigma_{dw}$, at a band junction between two domains and the deviation angle $\zeta$. The stress in the vicinity of the domain wall scales with the magnitude of the lattice distortion. The blue solid line represents the calculation based on Equation 3. The maximum stress in the vicinity of the domain wall increases with increasing deviation angle and c/a ratio. The c/a ratios of exemplary materials, including BT and tetragonal $Pb(Zr_{1-x}Ti_x)O_3$ with different titanium contents x are plotted for comparison as vertical lines. [208] e) Internal stress averaged over the whole polycrystalline body as a function of the c/a ratio for a La-doped PT. Average intrinsic stress increases with increasing the c/a ratio. TEM images of the domain structure are displayed for two different c/a ratios (indicated by the solid line). The TEM images display that the density of domain walls increases with increasing the c/a ratio. [214] Reprinted by permission from Springer Nature Customer Service Centre GmbH: [215], Copyright 2000.

4.1.2. Case studies

We next present two selected case studies, that allow to disentangle the complex interplay between crystal structure/lattice distortion and domain wall dynamics and shed light on functional properties.

4.1.2.1. Phase boundary in the $Pb(Zr,Ti)O_3$ system

Phase boundaries are important, because they are often associated with peaks in functional properties, such as piezoelectric or dielectric coefficients. [7] Various studies exist, which investigate the domain wall dynamics over phase boundaries, e.g., in PZT [100, 124, 218-223], $Ba(Zr_{0.2}Ti_{0.8})$-$(Ba_{0.7}Ca_{0.3})TiO_3$ [123, 125, 224], $(Bi_{1/2}Na_{1/2})TiO_3$-$(Bi_{1/2}K_{1/2})TiO_3$ [225], or $(Bi_{1/2}Na_{1/2})TiO_3$-$BaTiO_3$ [226, 227] systems. The most prominent phase boundary is most likely the rhombohedral-tetragonal phase boundary in PZT. In this section, we will discuss the domain wall dynamics in the rhombohedral (R) and tetragonal (T) phases in the PZT system, while we display the MPB composition for comparison. Multiple studies outlined that domain switching is easier in the rhombohedral phase compared to the tetragonal one, experimentally underlined by a higher activation field for polarization reversal [100, 124], a broader distribution of switching times [124], a lower Rayleigh coefficient, $\alpha$, [228, 229] a higher coercive field, [100, 221, 223] and a higher coercive stress [222, 230] for tetragonal PZT materials in comparison to rhombohedral counterparts.

While various reports are available on polycrystalline PZT, e.g., refs. [222, 231-233], we focus here on the $Pb_{0.985}V_{Pb_{0.005}}La_{0.01}(Zr_{1-x}Ti_x)O_3$ system, [234] since the domain-wall dynamics were characterized using a combination of time-dependent pulse and triangular bipolar HV electric measurements (Figure 6). [100] The polarization and strain hysteresis loops measured at 1 Hz and the time-dependent evolution of switched polarization and strain are displayed for different titanium contents $x$ in Figure 6a and b, respectively, facilitating a comparison between a T and R crystal symmetry. A MPB composition is displayed for comparison. The polarization and strain response of the R is sharper in comparison to the T compositions, which reflects the narrow distribution of the local electric fields in these materials (FWHM value in Table 1). [100, 235] The dependence of the activation barrier for polarization reversal (determined from time-dependent measurements), coercive field, bipolar strain, and remanent polarization are displayed as a function of the titanium content in Figure 6c. The impact of crystal structure on the switchable polarization can be quantified by the remanent polarization value. [100] A value of $P_r = 32.2\ \mu C/cm^2$ was found for the tetragonal composition ($x$=0.5), while a higher value was found for the rhombohedral one ($x$=0.4, $P_r = 37.6\ \mu C/cm^2$), reflecting the direct correlation to the calculated maximum switchable polarization fraction $P_{max}/P_s$, (Table 1). The coercive field and the activation barrier for polarization reversal (obtained from field- and time-dependent measurements [37]) in the T composition is by a factor of 2 higher compared to the R counterpart (Figure 6c), underlining that switching is more difficult in tetragonal materials. Particularly, it can be observed that the coercive field and the activation barrier scale with the distortion of the lattice, as displayed by the arrows in Figure 6c. The distortion of the tetragonal lattice for $x$=0.5, for example, is by a factor 3 higher compared to the rhombohedral material at $x$=0.4. [234] This outlines that domain wall movement gets more difficult with increasing lattice distortion, most likely related to enhanced stress in the vicinity of the interfaces, e.g., domain walls and grain boundaries, as discussed in section 4.1.1.2 and Figure 5.

The impact of the crystal structure on the fraction/dynamics of non-180° domain walls can be further quantified by the slope of the time-dependent strain curve (Figure 6b). The slope, which represents the fraction/dynamics of the moving non-180° domain walls (Figure 2) is lowest for $x$=0.5 (T). The movement of non-180° domain walls generates large internal stress in tetragonal materials [186], making the movement of these types of domain walls unlikely. The measurements provided in Figure 6b were corroborated by time-resolved synchrotron studies [27, 30, 89, 99, 236-238] and theoretical calculations [239]. While these studies confirm the movement of non-180° domain walls in tetragonal PZT compositions, their fraction is much smaller in comparison to R or MPB compositions. The contribution of non-180° switching to the volume fraction of domain switching processes was for example quantified as 20−25% for undoped rhombohedral PZT materials, while it was found to be lower for their tetragonal counterparts (7−8%). [240]

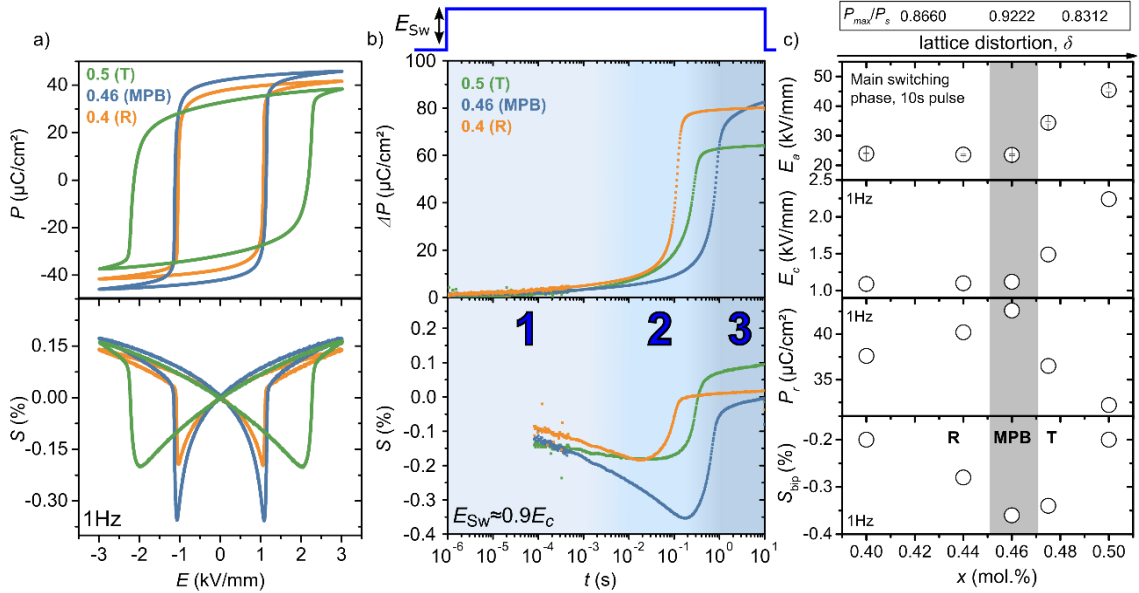

*Figure 6: Domain wall dynamics over the tetragonal-rhombohedral phase transition in a $Pb_{0.985}V_{Pb_{0.005}}La_{0.01}(Zr_{1-x}Ti_x)O_3$ model system. a) Macroscopic polarization and strain hysteresis loops measured at 1 Hz. A tetragonal, rhombohedral, and a composition with a phase mixture are compared. b) Representative dynamic curves of switched polarization, $\Delta P$, and macroscopic strain, S, of a poled sample as a response to a HV pulse of height $E_{Sw} \approx 0.9 E_C$. The numbers indicate the individual switching regimes according to ref. [89]. The dynamics of non-180° domain walls govern the first regime, while in the second regime the 180° switching events are dominant. Reprinted from [100], with the permission of AIP Publishing. c) Evolution of the activation barrier for polarization reversal of the main switching phase (determined from dynamic measurements [37]), the coercive field; the remanent polarization and the bipolar strain are displayed as a function of the composition. The change of the lattice distortion is indicated by the arrow. [100] The maximum possible fraction of single crystal polarization value, which can be achieved in a random polycrystalline state, $P_{max}/P_S$, for the different structures is indicated above. [181, 183, 184]*

4.1.2.2. Lattice distortion in La-doped $BiFeO_3$-$PbTiO_3$

The interplay between lattice distortion and domain wall dynamics was systematically studied for La doped tetragonal $BiFeO_3$-$PbTiO_3$ (BF-PT) [241-243], the tetragonal side of the BCZT system [244], lanthanum doped tetragonal PT [214, 215, 217], and KNN-based [245] materials. Exemplary results for BF-PT and BCZT are displayed in Figure 7. These studies outline that switching gets more difficult with increasing the $c/a$ ratio. At the same time, electromechanical properties were found to peak for small lattice distortions, followed by a decrease with increasing distortion away from the paraelectric phase. [232]

We will discuss the impact of lattice distortion in more details for the BF-PT and the BCZT systems, as the available data allows to establish a systematic link between the lattice distortion and functional properties. The impact of the $c/a$ ratio on the shape of the polarization hysteresis loops, as well as the ferroelectric and piezoelectric properties [241-243] are displayed in Figure 7a for the BF-PT system. With increasing the $c/a$ ratio, the polarization hysteresis loops broaden, and the ferroelectric behavior gets suppressed. The remanent and maximum polarization (measured at 8 kV/mm) decrease with increasing the $c/a$ ratio. For small lattice distortions, the $d_{33}$ response first peaks, followed by a decrease. The suppression of piezoelectric performance was explained by an incomplete poling of the material related to an increased distortion of the lattice. [241] As indicated by the coercive stress (Figure 7a), easiest switching was found for small $c/a$ ratios, while switching gets more difficult if the $c/a$ ratio increases. Moreover, a threshold value of $c/a_{thr}=1.045$ was reported, beyond which the ferroelectric/ferroelastic domain wall movement was suppressed under the electric or mechanical fields applied for the studied material. Diffraction measurements, provided in ref. [243], highlight the absence of change in the 002/200 peak intensities as a function of the applied mechanical stress, indicating the suppression of ferroelectric/ferroelastic domain wall movement. The mobility of the domain walls was further quantified by the Rayleigh parameters, $\varepsilon_0$, and $\alpha$. As displayed in Figure 7a, the Rayleigh parameters continuously decrease with increasing the $c/a$ ratio, indicating that reversible and irreversible domain wall motion gets reduced.

The impact of $c/a$ ratio on the functional properties of a tetragonal BCZT material is displayed in Figure 7b. [244] Macroscopic diffraction measurements show the poling process of two virgin samples with different $c/a$ ratios. As quantified by the change in 002/200 intensities, 90° domain reorientation was found to be easier for a lattice distortion of $c/a=1.0055$ compared to compositions with a larger lattice distortion of $c/a=1.0080$. Thereby, the contribution of 90° domain wall motion was found to continuously decrease with increasing the lattice distortion as quantified by the parameter $\Delta\eta/\Delta E$. Similar to polycrystalline BF-PT, the macroscopic $d_{33}$ in BCZT decreases, while the coercive field increases with increased lattice distortion. The continuous decrease of functional properties and the hindering of domain wall movement was related to the large internal stress, which builds up at the domain walls and grain boundaries, suppressing the ability of the polarization to align with the electric field, as mechanistically discussed in Figure 5.

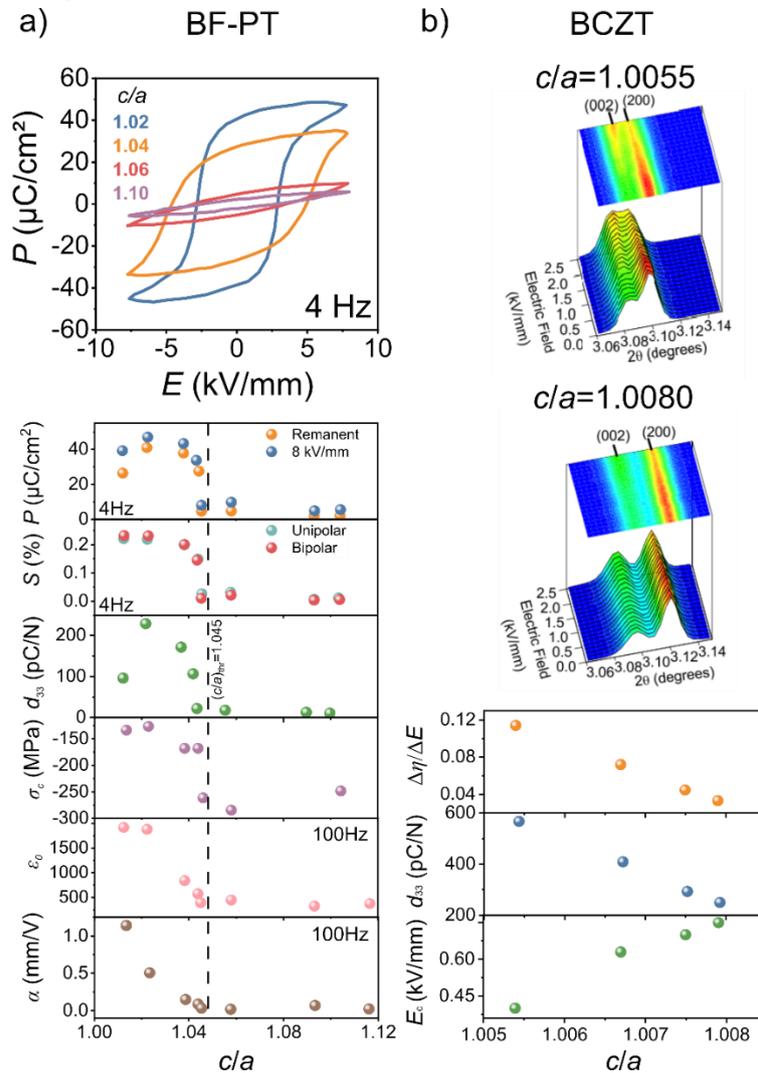

*Figure 7:* Impact of the $c/a$ ratio on domain wall dynamics and functional properties in polycrystalline ferroelectric/ferroelastic ceramics. a) The $c/a$ ratio was tailored by changing the La content in tetragonal BiFeO$_3$-PbTiO$_3$ polycrystalline ceramics. Field-dependent macroscopic polarization loops are displayed for different $c/a$ ratios. The polarization, P, macroscopic strain, S, macroscopic piezoelectric coefficient, $d_{33}$, coercive stress, $\sigma_c$, zero field permittivity, $\varepsilon_0$, and Rayleigh coefficient, $\alpha$, are displayed as a function of the $c/a$ ratio. A threshold value $c/a_{thr}=1.045$ was identified, beyond which polarization reversal was completely suppressed. [241-243] b) The $c/a$ ratio was changed through the composition in the tetragonal $(1-x)Ba(Zr_{0.2}Ti_{0.8})O_3-x(Ba_{0.7}Ca_{0.3})TiO_3$ system. Field-induced changes in the pseudo-cubic 002 diffraction peaks are displayed during poling from the virgin state for two different $c/a$ ratios. The change of the domain switching at weak electric fields ($\Delta E$ is from -0.45 kV/mm to +0.45 kV/mm), $\Delta\eta/\Delta E$, the macroscopic piezoelectric coefficient, $d_{33}$, and the coercive field are displayed as a function of the $c/a$ ratio. Reprinted from [244], with the permission of AIP Publishing.

4.1.3. Summary

Intrinsic structural parameters strongly impact the dynamics of ferroelectric/ferroelastic domain walls. In general, domain wall movement was found to be facilitated in materials with a higher number of available polarization directions and a smaller distortion of the lattice. Importantly, the easiness of domain wall movement, however, is not the only parameter impacting the electromechanical response of polycrystalline ferroelectric/ferroelastic materials. The maximum electromechanical response reflects the interplay between the lattice distortion and the mobility of non-180° domain walls. As outlined in section 4.1.2.1 and 4.1.2.2, these parameters are not independent of each other. A practical consequence of this interplay for example is, that the maximum electromechanical response can be found at the tetragonal side of the MPB in PZT materials. [232] Similar effects are observed in temperature-dependent studies of PZT [246, 247] and KNN-based [248] materials. Therefore, to optimize the electromechanical response of ferroelectric/ferroelastic materials, the crystal structure and the distortion of the lattice need to be tailored simultaneously.

Besides the dielectric and electromechanical properties discussed so far (Figure 6 and Figure 7), also other functional parameters can be engineered via crystal symmetry and lattice distortion. For example, the fracture toughness was found to be considerably reduced in tetragonal PZT compared to rhombohedral counterparts. [249] Similar to the piezoelectric response, the increase of fracture toughness due to the reorientation of ferroelastic domains, also referred to as the ferroelastic toughening effect, depends on the combined ability of the material for non-180° domain wall movement and the distortion of the unit cell. Furthermore, the mechanical quality factor, $Q$, which scales inversely with domain wall mobility in PZT, is considerably higher in tetragonal compared to rhombohedral compositions. [250]

4.2. Grain size and grain-orientation relationship

Representative hysteresis loops highlighting the impact of grain size [251] and grain-orientation [252] on domain wall motion are displayed in Figure 8 for polycrystalline BT. The polarization hysteresis loop of a BT single crystal is displayed in Figure 8b for comparison. [253] It can be observed that the polarization loops get slimmer if the microstructure of the material contains coarse grains or if the grains are highly aligned, implying facilitated domain wall movement for large grains and highly textured microstructures.

In comparison to previous reviews on the grain size-dependency [254-257] or the impact of crystallographic texturing [258-260] on dielectric and piezoelectric properties, we will focus on the domain wall-grain boundary interaction and will draw a link to the functional behavior. We will start by summarizing the impact of grain size on small- and large signal dielectric and electromechanical properties and discuss them in the framework of the static domain structure and the influence on electric and mechanical driving forces for domain wall dynamics (section 2.2.2). Next, we will apply these concepts in the context of textured polycrystals and outline how domain wall-grain boundary relationships can be used for functional property engineering.

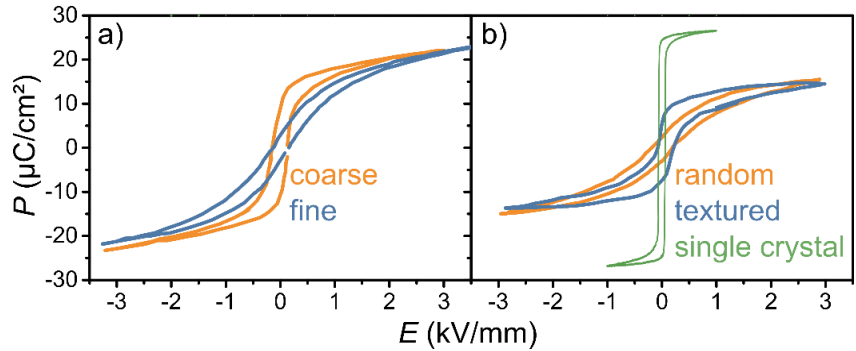

*Figure 8*: Impact of grain size and grain-orientation relationship on the shape of polarization loops. a) Polarization loops for polycrystalline BT ceramics with different grain sizes. [251] b) Polarization loops for random and crystallographically-textured polycrystalline BT ceramics. [252] The polarization loop of a BT single crystal is displayed for comparison. [253]

4.2.1. Grain size

4.2.1.1. Domain wall dynamics and domain morphology

The correlation between small/large signal properties and grain size is displayed in Figure 9a-d for polycrystalline BT [257, 261-280] and PZT [135, 139, 219, 281-291]. The relatively wide spread of absolute values may be related to different chemical compositions, defect states, crystal structures, utilized precursor materials or processing conditions, which will not be further discussed here. However, irrespectively of the spread of the data points, a general trend can be revealed. For both materials, the large signal properties (Figure 9a and b), i.e., the coercive field and the remanent polarization, are nearly independent of the grain size for large grains, while rapid changes occur if the grain size decreases. For small grain sizes, the coercive field increases, while the remanent polarization decreases with decreasing grain size, indicating the suppression of domain wall movement. Similar dependences can be observed in other ferroelectric/ferroelastic systems, such as KNN [292, 293], Ba(Zr$_{0.2}$Ti$_{0.8}$)O$_3$ [294] or BCZT [295], suggesting that the grain-size dependence of large signal properties is a general phenomenon.

The grain size-dependence of the small signal properties of BT and PZT is displayed in Figure 9c and Figure 9d, respectively. All studies report a peak in $\varepsilon_r$ and $d_{33}$ at ~1 µm for polycrystalline BT. For polycrystalline PZT, the reports on the dependence of the small signal properties on grain size, however, show much less unified characteristics. While some researchers find a continuous decrease of $\varepsilon_r$ [287], others observe a slight increase with decreasing grain size [139, 281, 284, 291].

To understand the correlation between grain size ($g$) and large/small signal properties, we display the domain size ($\delta$) as a function of grain size for BT [262, 264, 265, 267] and PZT [135, 288, 291] in Figure 9e and f, respectively. Two regions can be distinguished. The critical transition grain size, $g_{crit}$, between the two regions was calculated by Arlt considering elastic, electric, and surface energy of the domain walls. For polycrystalline BT, $g_{crit} \approx 5$ µm, while for PZT a value of $g_{crit} \approx 1-2$ µm was calculated. [251] Note, that the obtained experimental values of the critical transition grain size were found to be dependent on the chemical composition of the material. [291] Following the nomenclature of Arlt, [174] in the small grain size regime, a lamellar domain structure (contains only non-180° domain walls) is observed, while at large grain sizes, a banded domain structure (contains 180° and non-180° domain walls) develops. Domain images of a lamellar and a banded domain structure for BT [174] and PZT [291] are provided in Figure 9g and h, respectively.

The correlation between domain size and grain size is given by $\delta \propto g^m$. [296] For the investigated BT [262] and PZT [78, 291] materials $m \approx 0.5$ is typically observed, as displayed by the triangles in Figure 9e and f. The main driving force for the scaling behavior is the elastic strain, which is released via the formation of ferroelectric/ferroelastic domain walls. [174] A deviation from the scaling behavior can be observed when the grain size is decreased, e.g., for $g<0.5$ µm in BT (Figure 9e). At even smaller grain sizes (a theoretically calculated value for BT is around 40 nm [174]) ferroelectricity vanishes as experimentally proved for both BT and PZT. [287, 294, 297-299]

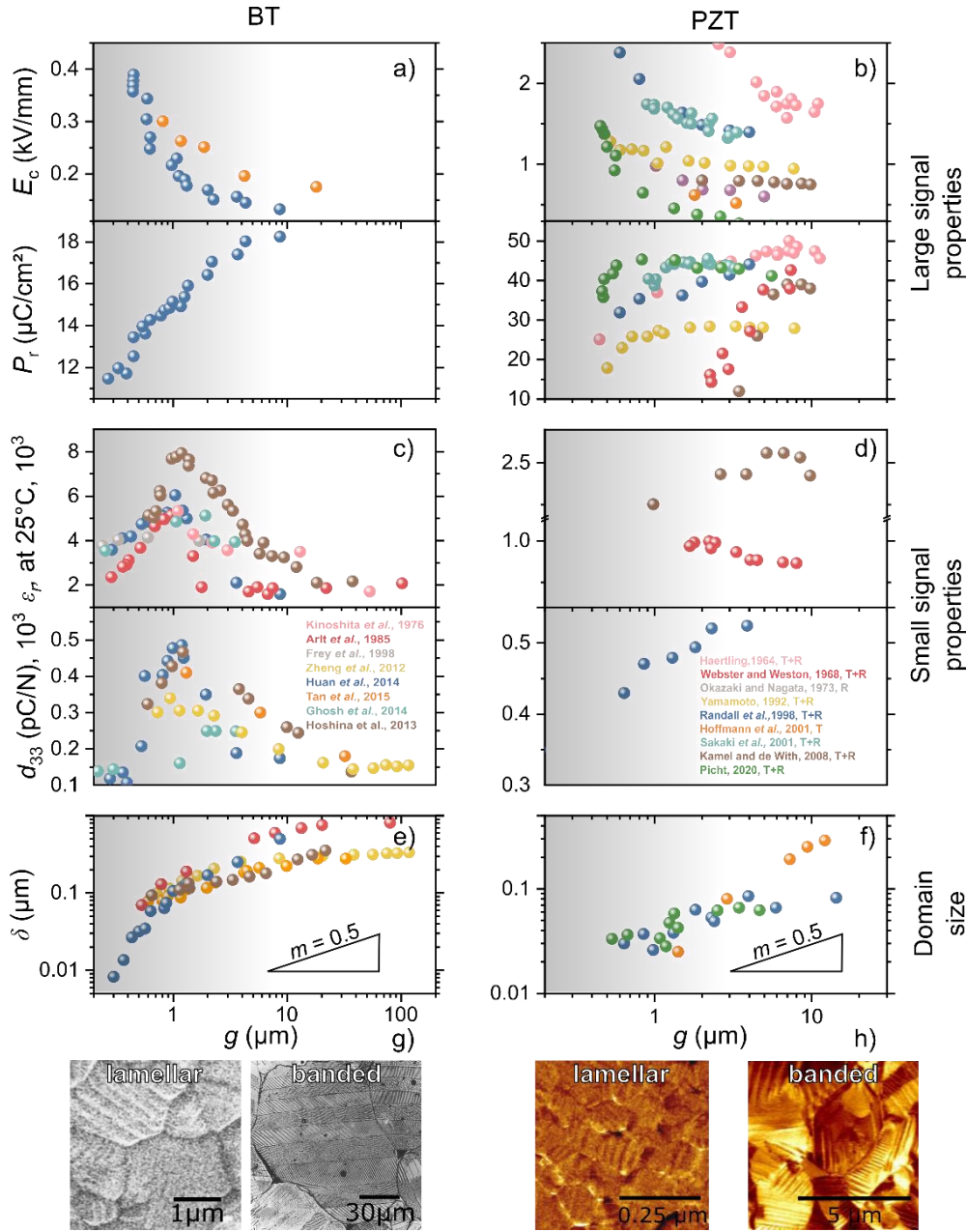

*Figure 9:* *Impact of grain size on ferroelectric and electromechanical properties, as well as domain sizes for polycrystalline BT and PZT. The dependence of the a, b) large signal properties (coercive field and remanent polarization) and c, d) small signal properties (permittivity and piezoelectric coefficient) is displayed together with the domain size in e) and f). The triangles in e) and f) indicate the typical correlation between grain and domain size, $\delta \propto g^m$, with $m \approx 0.5$, [296] for BT [257, 261-268] and PZT [135, 283-291]. For PZT the crystal structure of the composition is abbreviated with T (tetragonal) and R (rhombohedral). [174] Microscopy images for a lamellar and banded domain structure of polycrystalline BT [174] and of PZT ceramics [291] are displayed in g) and h). g) Reprinted by permission from Springer Nature Customer Service Centre GmbH: [174], Copyright 1990. h) Reprinted from [291], with the permission of AIP Publishing.*

4.2.1.2.   Grain size dependence of domain wall-related contributions under large and small electric fields

Under large electric fields (Figure 9a and b), the contribution of non-180° domain wall movement in polycrystalline ferroelectric/ferroelastic materials decreases with decreasing grain size. For PZT, XRD measurements revealed that non-180° domain wall motion is considerably reduced in fine (~1.6 µm) compared to coarse grained (~3.3 µm) ceramics. [288] Similar conclusions were drawn for tetragonal $Pb_{0.9}La_{0.1}TiO_3$ in the grain size range of 1.3–6.0 µm. [300] For polycrystalline BT the domain wall movement is significantly higher for grain sizes of 2.0 µm and 3.5 µm in comparison to counterparts with a smaller grain size of 0.3 µm. [268] In analogy to bulk polycrystalline ceramics, non-180° domain switching was found to be negligible in polycrystalline thin films with a grain size smaller than 2 µm, while it was evident for larger (>5 µm) grain sizes. [301]

While domain wall mobility decreases with decreasing grain size under large electric fields, the domain-wall related response under small electric fields is considerably different. The peak in the small signal relative permittivity and piezoelectric coefficient of BT, which is observed for a grain size of ~1 µm (Figure 9c) was explained by enhanced domain-wall-related contributions. [255, 256] In this frame, the grain-size dependent small signal response was quantified for BT using XRD [268] and spectroscopic techniques [257, 273], as summarized in Figure 10. The domain switching fraction, $\eta_{002}$, obtained from XRD is displayed as a function of grain size for BT in Figure 10a. For subcoercive fields (40% of the coercive field) a higher degree of non-180° domain switching fraction ($\Delta\eta = 0.013 \pm 0.02$) was found for materials with a grain size, $g$=1.97±0.44 µm, while the sample with a large grain size, $g$=3.52±0.29 µm, featured a smaller non-180° domain switching fraction of $\Delta\eta$=0.007±0.002. [268] Complementary to XRD measurements, frequency-dependent small-signal measurements facilitate the distinction between intrinsic and extrinsic contributions. The frequency-dependent permittivity up to the terahertz range for polycrystalline BT of different grain sizes is displayed in Figure 10b. [273] Extrinsic domain wall related contributions feature a characteristic relaxation frequency at $f$~$10^8$ Hz, while intrinsic contributions vanish at frequencies higher than $f$~$10^{11}$ Hz. The determined domain-wall related contribution to the permittivity, $\varepsilon_r^{\text{ext}}$, are displayed as a function of grain size in the inset of Figure 10b. A peak contribution was found for a grain size of 1.4 µm, while substantially lower values were reported for BT materials with larger (13.0 µm) and smaller (0.7 µm) grain sizes.

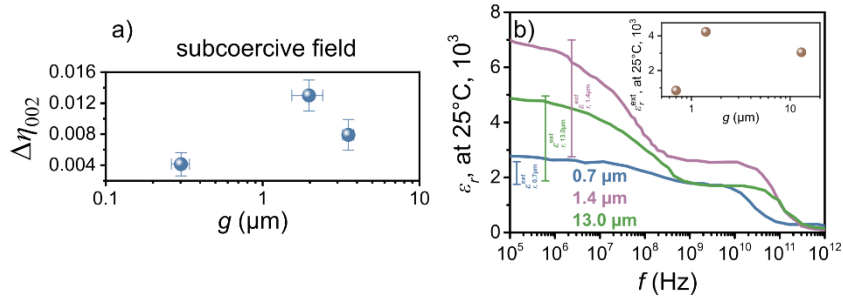

*Figure 10: Impact of grain size on extrinsic contributions in polycrystalline materials under small electric fields. a) Domain switching as a function of grain size during subcoercive electric field application in polycrystalline BT (applied E=±0.2kV/mm, E$_c$=0.5kV/mm). [268]. b) Frequency dependence of the dielectric constant for polycrystalline BT materials with different grain sizes. The grain-size dependence of the extrinsic contribution is quantified in the inset. [273]*

4.2.1.3.   Interplay between elastic strains and domain wall dynamics in the vicinity of grain boundaries

In order to understand why the impact of grain size on the large and small signal response is different, we continue with an explanation of the interplay between domain wall movement, elastic strains, and grain size. The impact of grain size on lattice strains (quantified by the *c/a* ratio) is displayed for polycrystalline BT with different grain sizes in Figure 11a. Above a grain size of ~1.5 µm, lattice distortion was found to be independent of the grain size in polycrystalline BT, [262, 302] while below ~1.5 µm, the lattice distortion continuously drops. A similar behavior is found in polycrystalline PZT and composition-dependent critical grain sizes in the range of 0.6 µm [291] to 1.5 µm [282] were reported. In polycrystalline BT the drop in the lattice distortion comes along with a decrease of the Curie temperature, corroborating the decreased stability of the tetragonal phase. As displayed in Figure 11b, the Curie temperature rapidly decreases when the grain size is reduced into the submicrometer range, [266, 274] while it remains unchanged for larger grain sizes [261, 276, 303, 304]. At the same time, the orthorhombic-rhombohedral phase transition temperature slightly increases with decreasing grain size. [264, 305] Interestingly, and in comparison to the discussed impact of the *c/a* ratio (section 4.1.2.2), the coercive field and remanent polarization do not decrease with decreasing grain size (Figure 9a and b), indicating that the decreasing *c/a* ratio is not the dominant mechanism.

The impact of grain size on microstrains was quantified by the broadness of the diffraction peaks in XRD measurements. [306] The broadness of the (222) peak, $\beta$, which is proportional to the microstrains, increases with decreasing grain size, as displayed for polycrystalline BT materials over a grain size range of 0.15—50 µm (Figure 11c). [307] A qualitatively similar behavior was found for a polycrystalline Pb(Zr$_{0.7}$Ti$_{0.3}$)O$_3$ over a grain size range of 3.9—10.4 µm. [139] An additional feature, which is frequently revealed by XRD measurements, is the occurrence of diffuse scattering between the tetragonal (002) and (200) peaks [264, 268-270] when the grain size is decreased. Such features were observed for BT [194, 195, 262, 270] and PZT [193] polycrystals and explained by microstrains located in the vicinity of ferroelectric/ferroelastic domain walls.

While it is clear that the lattice and microstrains decrease with decreasing grain size, maps of the resulting strain state were obtained spatially resolved. Strained regions were found to be located in the vicinity of domain walls [170, 308, 309] or grain boundaries [171, 310, 311]. The strain profile is displayed as a function of the distance from the grain boundary in Figure 11d for a grain with a diameter of 30 µm embedded in a polycrystalline BCZT matrix. The strains in the vicinity of the grain boundary are about twice as high as the strains in the center of the grain. [209, 210, 312] Together with the spatial variation of strain, the spatial variation of the density of domain variants, $\delta$, is displayed in Figure 11d. As displayed by the red shaded area, both parameters peak in the vicinity (2-3 µm) of the grain boundary, and the domain density is enhanced by 30% in comparison to deep inside the grain, [171] related to the strain releasing effects of ferroelectric/ferroelastic domain walls. The findings displayed in Figure 11d agree well with bulk polycrystalline [139, 313] and thin film PZT materials [314], for which the domain wall density was found to be substantially enhanced in the vicinity of a grain boundary.

The domain wall density directly impacts the mobility of the domain walls and calculations predict a reduced domain wall mobility in finer domain structures than in courser ones. [206, 262, 315] Figure 11e displays a PFM image of a grain of a polycrystalline Pb(Zr$_{0.7}$Ti$_{0.3}$)O$_3$ ceramic with the grain boundary highlighted as a red line, [139] facilitating to quantify coercive voltages spatially resolved. The coercive voltage map highlight that domain wall dynamics are hindered in volumes in the vicinity of the grain boundary, in particular in areas with high domain wall density. The findings on polycrystalline bulk materials are in good agreement to reports on polycrystalline PZT thin films (Figure 11f), where domain wall motion was found to be impacted in a range of 450±30 nm in the vicinity of a 24° tilt grain boundary in a rhombohedral Pb(Zr$_{0.52}$Ti$_{0.48}$)O$_3$ bicrystal. [316] Maps of the nonlinear domain-wall related response along the grain boundary are displayed in Figure 11f and a significant dip can be found at the position of the grain boundary. Furthermore, local hysteresis loops measured at the grain boundary were reported to have a considerably larger coercive field, [134, 317] display a strong imprint, [132, 318] and incomplete domain reorientation even under high electric fields. [319] In addition, domain wall pinning at the grain boundary is frequently reported experimentally [318, 320-322] and by simulations [323]. The volume influenced in the vicinity of the grain boundary thereby can be tuned by the composition of the material, boundary orientation angle, as well as temperature. [141, 314, 316] Domain wall continuity over grain boundaries, as experimentally demonstrated for grain boundaries in polycrystalline BT [324, 325] and PZT [140, 326, 327] may further facilitate domain wall dynamics for specific grain configurations. For example, one of the grain boundaries that allows domain continuity in tetragonal materials is the 111 boundary, allowing 110 plane matching [328, 329]. For more insights into detailed grain configurations, we refer the reader to refs. [330, 331] and a review article (ref. [332]).

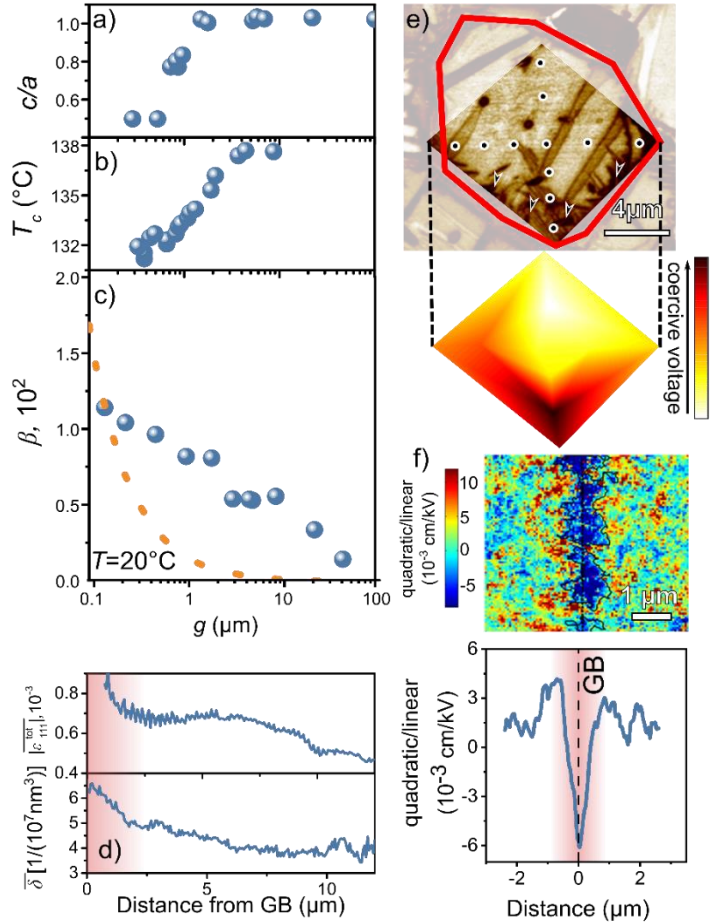

*Figure 11: Correlations between elastic strain and ferroelectric/ferroelastic domain walls. Variation of the a) the lattice distortion, [262] b) the Curie temperature [266] and c) the broadness, $\beta$, of the (222) diffraction peak, [307] as a function of grain size of polycrystalline BT. Contributions of the crystallite size (Scherrer broadening) to the broadening of the diffraction peaks are indicated by the orange line and cannot explain the peak broadening alone. d) Dependence of the total strain, $\varepsilon_{111}^{tot}$, and the density of domain variants, $\delta$, measured as a function of the distance of the grain boundary of a slice of a 30 μm large grain for a BCZT polycrystal in the virgin state. Data is taken from ref. [171]. The red shaded area indicates the impact of the grain boundary on the simultaneous enhancement of the local strains and domain wall density. e) PFM image of a grain of a polycrystalline rhombohedral PZT ceramic. The red line indicates the position of the grain boundary. Areas of high domain wall density are indicated by arrows. Local coercive voltages were determined by local switching measurements (measurement positions are indicated by dots) and are displayed as a contour plot for the same grain region. Reprinted from [139], with the permission from Elsevier. f) Maps of the nonlinear response measured across a grain boundary of a $Pb(Zr_{0.45}Ti_{0.55})O_3$ thin film. The dashed line indicates the position of the grain boundary, while the area shaded in red indicates the substantial impact of the grain boundary on domain-wall mobility. Reproduced from [314], with the permission of Wiley.*

#### 4.2.1.4. Summary

The impact of grain size has been intensively studied for polycrystalline BT and PZT. For both materials, the large-signal coercive field increases, while the remanent polarization and the non-180° domain switching fraction continuously decrease with decreasing grain size (Figure 9a and b). This effect was related to strained volumes in the micrometer vicinity of the grain boundary, which comes along with a considerably enhanced domain wall density (Figure 11d). [171] As a consequence, domain wall movement gets locally reduced in the vicinity of the grain boundary (Figure 11e and f). [139, 314] For grain sizes below the critical transition grain size, discussed in section 4.2.1.1, the inhomogeneous stress state occupies the entire grain. This is reflected in a rapid decrease in the lattice distortion (Figure 11a, [262]) and the development of a lamellar domain structure (Figure 9g and h), which nearly completely suppresses domain wall movement under large electric signals. [333] Macroscopically, the coercive field and remanent polarization significantly change in the small grain size regime (Figure 9a and b). In order to obtain enhanced large-signal electromechanical properties, polycrystalline materials with large grain sizes featuring a limited amount of grain boundaries are therefore desirable. In applications of ferroelectrics/ferroelastics in multilayer actuators, however, additional design criteria, such as the thickness of each layer needs to be considered to avoid leakage currents or device failures. [334] Going beyond the discussed impact on the electromechanical response, also fracture toughness, [335, 336] electric fatigue, [311, 337] and high-power behavior [289] can be tailored by grain size engineering.

The dependence of small-signal properties on the grain size is more complex (Figure 9c and d). With decreasing grain size, the non-180° domain density is considerably enhanced (Figure 9e and f), and particularly regions in the vicinity of the grain boundary are impacted (Figure 11d and e). Vibrational motion of these non-180° domain walls under low voltages are the main reason for the enhancement of the small signal dielectric permittivity and piezoelectric response (Figure 10a and b). [268, 273] The decrease of the *c/a* ratio (Figure 11a) may further enhance domain wall mobility, as discussed in section 4.1.2.2. Since the grain-size-dependence of small signal properties is temperature-independent, [251] grain size induced structural changes, e.g., an orthorhombic/tetragonal phase coexistence, [280, 305] originating from the grain-size-dependency of the phase transition temperatures contribute as a secondary effect to the enhancement of the small signal properties. Decreasing the grain size further below 1 μm leads to a rapidly increasing mechanical stress (Figure 11a-c) and a suppression of the domain wall movement, which comes along with a decrease of dielectric and piezoelectric performance and the subsequent loss of ferroelectricity in the sub-100 nm range. While the enhancement of the small signal properties naturally occurs in polycrystalline BT materials at a grain size of ~1 μm, a similar, but much less emphasized behavior was observed for polycrystalline PZT materials. [139, 284, 291] For example, the solid solution Pb(Zr,Ti)$O_3$-Sr($K_{0.25}$$Nb_{0.75}$)$O_3$ displays a peak of the dielectric permittivity in the grain size range of 2–3 μm. [281]

4.2.2. Crystallographic texture

As we have outlined in section 4.2.1, elastic strains substantially influence domain wall dynamics in polycrystalline ferroelectric/ferroelastic ceramics with a random grain orientation. Crystallographic texturing impacts elastic strains, as reported for structural [210, 338, 339] and functional [132, 340, 341] polycrstals. To visualize this effect, we display representative stress distributions comparing a random and a highly textured $Al_2O_3$ in Figure 12a. [210] The distribution is narrow for a textured material and the absolute value of stress is substantially lower in comparison to a non-textured counterpart, which was explained by lower grain boundary misorientation angles and energies. [210] Model experiments on thin films [141] supported by phase field simulations [342] have investigated the resulting impact on domain wall dynamics, suggesting a pronounced dependency on the misorientation angle. A summary of this effect is provided in Figure 12b and c. Figure 12b provides a schematic drawing of two grains highlighting the tilt and twist angle, while the impact on the width of the region with reduced domain wall mobility is outlined in Figure 12c for a rhombohedral Pb($Zr_{0.52}Ti_{0.48}$)$O_3$ and a tetragonal Pb($Zr_{0.45}Ti_{0.55}$)$O_3$ composition. [141] For small misorientation angles, typically found in textured polycrystalline materials, the width of the region with reduced nonlinear response is nearly zero and domain wall mobility is not impacted. With increasing misorientation angle, the width of the region continuously increases to ~0.3 and ~0.6 μm for a tilt and twist grain boundary, respectively, indicating a more pronounced impact on domain wall dynamics (Figure 12c).

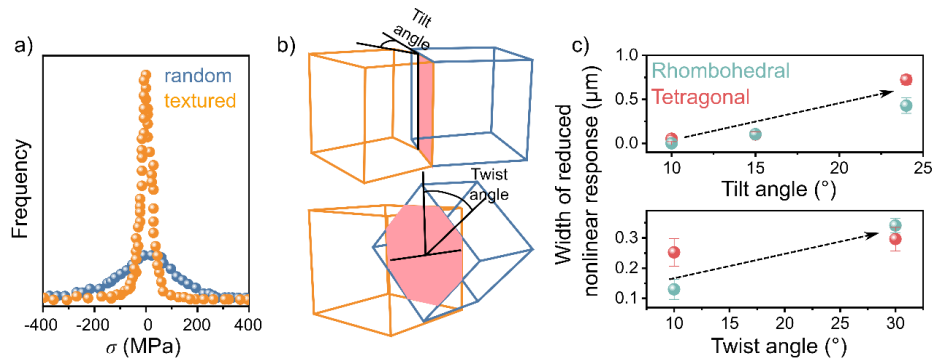

*Figure 12:* Impact of crystallographic texture on the stress and domain wall mobility. a) Residual stress distribution in textured and random polycrystalline $Al_2O_3$. The stress is significantly reduced in a polycrystalline ceramic with textured grains compared to a random counterpart. [210] b) Schematic representation of a tilt and a twist grain boundary separating two grains. The grain boundary area between two grains is indicated by the red area. The tilt and twist angles are schematically displayed. c) Experimental results summarizing the impact of twist and tilt angle on the width of the region with reduced nonlinear response in the vicinity of a grain boundary for a rhombohedral Pb($Zr_{0.52}Ti_{0.48}$)$O_3$ and tetragonal Pb($Zr_{0.45}Ti_{0.55}$)$O_3$ bicrystal model system. As indicated by the dashed arrows, the width of the region increases with increasing tilt/twist angle for both crystal structures. [141]

Literature on the impact of crystallographic texture on the easiness of domain wall dynamics, however, is inconsistent. Some reports find that crystallographic texturing facilitates domain wall movement [252, 341, 343], while others find hindering effects [344-346]. Again, other report that domain wall dynamics are not impacted by crystallographic texturing [347-349]. To disentangle different experimental results, we will review the impact of orientation of the grains with respect to the electric field (section 4.2.2.1), the degree of crystallographic orientation (section 4.2.2.2), and the impact of reactive templates (section 4.2.2.3) on the domain wall mobility and functional performance in the following.

4.2.2.1. Anisotropy

The domain wall dynamics depend on the orientation of the crystallographic texture with respect to the external electric field. This was demonstrated exemplarily for an Aurivillius oxide bismuth titanate ($Na_{0.5}Bi_{4.5}Ti_4O_{15}$) model systems, which has a highly anisotropic unit cell (the c axis is 7.5 times the a/b axis [350]) and allow 180° and 90° domain switching. [351, 352] Polarization hysteresis loops of samples with grains oriented parallel and perpendicular to the c-axis are compared to a random counterpart in Figure 13a. A high remanent polarization of 21 μC/cm² was found if the electric field was applied parallel to the c-axis, while the domain wall motion was completely suppressed for electric fields perpendicular to it. The enhanced switchable polarization for the sample with polarization parallel to the c-axis in comparison to the random counterpart was attributed to a synergistic alignment of the ferroelectric distortion directions in each grain during domain switching. [351] In analogy to clamping effects due to in-plane stress in thin films, [353, 354] the absence of domain wall movement was related to the highly anisotropic unit cell, preventing reorientation. Similar results were observed for textured $Sr_{0.53}Ba_{0.47}Nb_2O_6$ (SBN) [355] and KNN [356], indicating that domain wall movement is easiest if the long axis is oriented parallel to the applied electric field.

4.2.2.2. Degree of crystallographic texture

The impact of the degree of crystallographic texture was systematically investigated on a series of ($Ba_{0.85}Ca_{0.15}$)$TiO_3$ (BCT) samples with different degrees of crystallographic texture in refs. [126, 341]. The data are displayed in Figure 13b-d. Figure 13b displays polarization loops of samples with different degrees of crystallographic texture, quantified by the Lotgering factor, $F$. [357] A highly textured sample is characterized by $F$ with values close to 100%, while a sample with random grain orientation has a value of $F$=0%. For the BCT series, with increasing $F$, the squareness of the polarization loops increases indicating enhanced and facilitated domain wall dynamics. Similar behavior was also observed in KNN-based materials with a high degree of crystallographic texture [358, 359], textured SBN [355], as well as in textured thin films [360, 361].

Further information on the impact of crystallographic texture on the dynamics of domain walls in polycrystalline BCT was obtained from time-dependent polarization switching measurements. Figure 13c displays the time dependence of the switched polarization of the BCT samples with different degrees of crystallographic orientation under an applied electric field pulse of height $E_{Sw}$ = 0.6 kV/mm. A (001)-oriented BT single crystal is displayed for comparison. With increasing the degree of crystallographic texture, the switching time of the sample with a high degree of crystallographic texture ($F$=83%) is 2–3 orders of magnitude smaller in comparison to a sample with a low degree of crystallographic texture ($F$=26%). The response of the polycrystalline samples, however, remain far behind the BT single crystal. The impact of the degree of crystallographic texture on domain wall dynamics was quantified by the coercive field and the activation barrier for domain wall movement. These parameters



are displayed as a function of the Lotgering factor in Figure 13d. The coercive field of the highly textured samples (0.55 kV/mm) is about 18% lower compared to counterparts with a low degree of crystallographic texture (0.60-0.67 kV/mm). For comparison the coercive field of a BT single crystal was found to be about 0.05 kV/mm [362], which is about an order of magnitude lower. Similar as the coercive field, also the activation barrier for polarization reversal decreases with increasing Lotgering factor (Figure 13d).

As outlined in Figure 12a, the internal stress decrease with increasing the degree of crystallographic texture, allowing a facilitated cooperative alignment of the polarization vectors during the application of an electric field. For the series of BCT, reduced internal stress can be manifested by an increased *c/a* ratio, which comes along with an increased Curie temperature, as indicated by the arrow in Figure 13d. As discussed in section 4.1.2.2, high lattice distortion in polycrystalline materials with a random orientation of grains hinders domain wall movement. However, as displayed in Figure 13d, this effect is different, if the microstructure is textured and a high lattice distortion in combination with a high degree of crystallographic texture was found to be beneficial for domain wall dynamics.

### 4.2.2.3. Non-reactive templates

Beyond the orientation relationship and the degree of crystallographic texture discussed so far, the interactions between the templates and the ferroelectric matrix are important if non-reactive templates are utilized, i.e., templates that remain present in the matrix after sintering. This impact of the non-reactive templates on domain wall dynamics and functional properties was studied systematically in $Pb(In_{1/2}Nb_{1/2})O_3$-$Pb(Mg_{1/3}Nb_{2/3})O_3$-$PbTiO_3$ polycrystals, which were textured by BT templates (Figure 13e and f). [347] Figure 13e displays the impact of the volume fraction of BT templates on the Lotgering factor, the coercive field and remanent polarization. It was found that the Lotgering factor continuously increases with increasing BT concentration. At the same time the coercive field increases, and the remanent polarization drops, if the concentration of BT templates exceeds 1 vol.%. This indicates that with increasing BT template concentration, domain wall motion gets more difficult, even though the degree of crystallographic texture is increasing, which is different to the behavior discussed in section 4.2.2.2.

To understand the different behavior, the microstructure containing the BT templates is displayed in Figure 13f. [347] In comparison to the results presented in section 4.2.2.2, where reactive templates were used to fabricate the textured polycrystals, [363] BT templates are not reacting with the matrix material. This impacts the distribution of elastic stress in the matrix and thus the mobility of the domain walls. Using finite element simulations, local stress up to 8 MPa was calculated, located in the vicinity of an $SrTiO_3$ template in a $Pb(Mg_{1/3}Nb_{2/3})O_3$-$PbTiO_3$ (PMN-PT) matrix. [259] These elastic stresses were further found to result in a self-poling effect, i.e., the development of a preferential ferroelastic texture already during processing. [364]

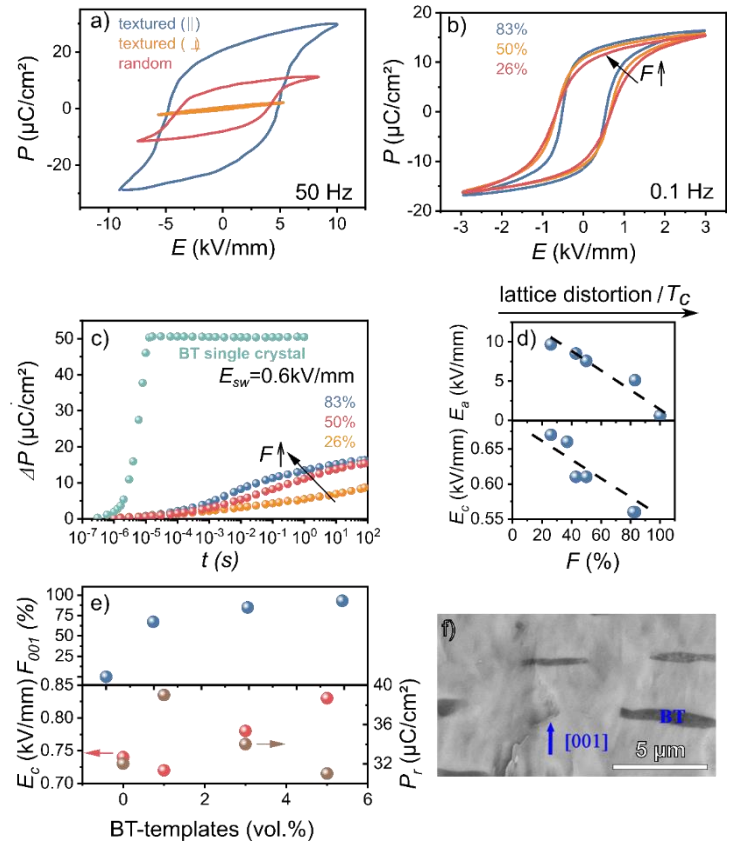

*Figure 13: Experimental results on the impact of crystallographic texture on domain wall dynamics: a) Impact of the orientation of texture on the polarization hysteresis loops (measured at 200°C) in textured $Na_{0.5}Bi_{4.5}Ti_4O_{15}$. The orientation of the crystallographic texture is parallel and perpendicular to the electric field. A polycrystalline material with random grain orientation is displayed for comparison. [352] b) Large signal polarization and strain loops for polycrystalline BCT with different degrees of crystallographic texture (the degree of texture is quantified by the Lotgering factor, F [357]). [341] c) Polarization dynamics for polycrystalline BCT with different degrees of crystallographic texture (measured under an electric field pulse of height $E_{Sw}$=0.6kV/mm). A BT single crystal is displayed for comparison. [126] d) The coercive field [341] and the activation barrier [126] for polarization reversal for polycrystalline BCT are displayed as a function of the Lotgering factor. The change of the lattice distortion and the Curie temperature is indicated by the arrow. e) The impact of the amount of BT templates in polycrystalline PMT-PT on the Lotgering factor, $F_{001}$, is displayed. The effects on the coercive field, and the remanent polarization are highlighted as well. A micrograph of the material with 5 vol.% BT templates is displayed in f). Reprinted from [347], with the permission of AIP Publishing.*



4.2.2.4. Summary

Crystallographic texturing enables tailoring of the dynamics of domain walls as a consequence of a stress reduction (Figure 12a), which facilitates the domain wall movement. A side effect, coming along with the stress reduction during the electric field-application is an enhancement of the dielectric breakdown field of textured materials in comparison to random counterparts. [365] Crystallographic texturing is currently applied to make full use of the intrinsic dielectric and piezoelectric properties, mimicking the behavior of single crystals. Even though domain wall dynamics are facilitated, experiments [366] and simulations [367] find only small contributions of extrinsic ferroelectric/ferroelastic domain wall motion to small signal properties in poled textured ceramics. Phase field simulations, for example, suggest that domain walls contribute about 50% to the small signal piezoelectric and permittivity in randomly oriented polycrystals, while the contribution was found to be negligible in textured counterparts. [367] To exploit the domain-wall related contributions to the dielectric and piezoelectric performance in textured polycrystals further, future studies should optimize the angle between the applied electric field and the orientation of the crystallographic texture. Furthermore, studies on the impact of crystallographic texture on the ferroelectric domain structure via domain imaging techniques are desirable.

For single crystalline materials it is known that the small signal piezoelectric coefficient increases with decreasing domain size. [368, 369] Building up on this, a strategy was suggested to combine crystallographic texturing and domain engineering in polycrystalline ceramics to improve their small signal response. This was demonstrated for two textured BT materials with approximately the same degree of crystallographic texturing ($F\sim83\%$) and different grain sizes. [370] For a grain size of 1.2 µm a piezoelectric response of 507 pC/N was reported, while a higher value of 788 pC/N was found for a reduced grains size of 0.8 µm. Interestingly, this value is substantially higher than the reported values for all grain sizes in Figure 9a, indicating the potential for an additional boost in functional properties if grain size and crystallographic texture are combined. Colossal piezoelectric coefficients of up to 10000 pC/N were foreshadowed using a synergetic approach of domain engineering and crystallographic texturing. [370]

4.3. Porosity

In comparison to grain size (section 4.2.1) and crystallographic texture (section 4.2.2), porosity relieves mechanical constraints and broadens the distribution of the local electric field. The latter effect is mainly related to a large difference in the relative permittivity between the ferroelectric matrix (e.g., $\varepsilon_r = 1500$ for PZT [101]) and the pores ($\varepsilon_r = 1$ for air). This effect is particularly enhanced for regions in the vicinity of the pores. [371] Typically, finite element methods (FEM) facilitate calculation of the distribution of the local electric field. A comprehensive FEM study of the impact of porosity on the distribution of the local electric field is displayed in Figure 14 for different degrees of porosity and pore morphologies. [128] Here, $E_{app}$ is the externally applied electric field to the porous body and the simulated distribution of the local electric field, $E_{loc}$, is displayed. Independent of the pore morphology, the presence of porosity shifts the distribution of the local electric field towards lower values and broadens it. The amount, shape, and orientation of the pores are levers to tune the local electric field distribution.

Figure 14a displays the impact of total porosity (5–27 vol.%) on the local electric field distribution. The simulations reveal a broadening of the distribution with increasing degree of porosity. [372-374] The impact of the radius of isometric pore size under a constant total porosity on the distribution of the local electric field is displayed in Figure 14b. For pore radii in the range of 20−120 µm, the local electric field distribution was found to be unaffected by the pore radius. Complementary studies find an impact of pore diameter on the distribution of the local electric field, if the pore radius is extended towards the nanometer range. [375] Finally, the interplay between the orientation angle of anisometric pores and the distribution of the local electric field is displayed in Figure 14c. With increasing pore orientation angle, the distribution of the local electric field shifts towards lower electric fields and broadens. [101, 128, 371, 373, 376]

Besides the local electric driving forces discussed so far, porosity also impacts local mechanical driving forces for domain wall motion. The domain switching fraction of ferroelectric/ferroelastic domains was enhanced in porous PZT thin films [377-379] compared to dense counterparts, which was related to declamping effects in the vicinity of the pores. [380-382] Corroborating this mechanism, theoretical results outline facilitated domain switching in the vicinity of surfaces. [383] According to a synchrotron study on porous BaTiO$_3$, residual stress decreases from 70 MPa to 40 MPa with increasing porosity, [384] indicating that porosity reduces intergranular stress in ceramic polycrystals. Such declamping of individual grains was demonstrated to result in giant functional properties, e.g., piezoelectric response and energy-harvesting figures of merit. [385]

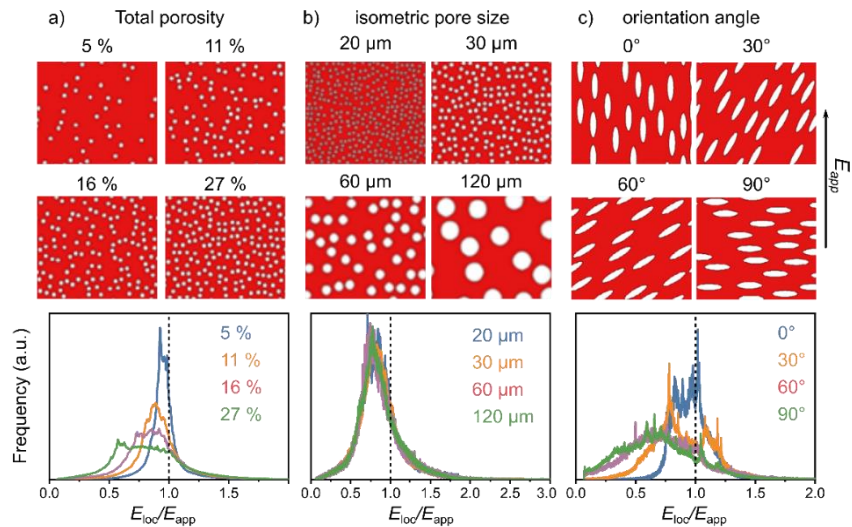

*Figure 14:* The impact of pores on the statistical distribution of the local electric field, $E_{loc}$. The pore morphologies consider changes in a) total porosity, b) isometric pore size, and c) orientation angle with respect to the electric field, $E_{app}$. The 2D FEM simulations take the impact of the porosity on the distribution of the local electric field into account. [122] For a completely dense material, the local electric field would be equal to the applied electric field as displayed by the dashed line.



4.3.1. Degree of total isometric porosity

To discuss the impact of total isometric porosity on domain wall dynamics and functional properties, we present experimental results on samples with different degree of total isometric porosity. Exemplary SEM micrographs of a dense and a porous (degree of porosity, $p$ = 40%) BCZT (0.5Ba(Ca$_{0.8}$Zr$_{0.2}$)O$_3$-0.5(Ba$_{0.7}$Ca$_{0.3}$)TiO$_3$) material are compared in Figure 15a. Polarization loops of the same material with different degrees of isometric porosity are displayed in Figure 15b. [374] With increasing the degree of porosity, the remanent and saturated polarization continuously decrease, while at the same time the polarization loops broaden. Similar results have been reported for porous PZT [128, 372, 386-389] and other BCZT [390] compositions. The impact of porosity on the dynamics of domain walls during polarization reversal was studied for a dense and a porous ($p$=40%) PZT material and the switched polarization as a function of time is displayed in Figure 15c. [128] The results indicate that the switched polarization and the dynamic of domain walls get reduced for the material with isometric pores. The latter is reflected by the decreased slope at the inflection point of the polarization curves, as highlighted in Figure 15c. The shape of the field- and time dependent polarization response in porous materials (Figure 15b and c) was explained by the distribution of the local electric field, which shifts to lower values and broadens with increasing the degree of isometric porosity, as displayed in Figure 14a. [128]

An additional effect observed in the polarization hysteresis loops in Figure 15b is the reduction in the remanent polarization with increasing volume fraction of isometric porosity. The fraction of the remanent polarization of a porous material, $P_r$, normalized to the remanent polarization of a dense counterpart, $P_r^0$, is displayed for different materials in Figure 15d. [372, 374, 387] The decrease of $P_r/P_r^0$ was interpreted as a reduced degree of poling and is responsible for the reduced piezoelectric coefficient of porous materials compared to dense counterparts. [391-393] If the entire porous material would be polarized, the fraction $P_r/P_r^0$ should follow the dashed line, as displayed in Figure 15d. The experimentally measured fraction of porous materials, however, is significantly lower, as indicated by the solid line. The difference indicates that a substantial amount of the material remains unpoled after the application of the electric field. The difference between the experimentally observed and the theoretical limit is considered mathematically by the depolarization factor, $d_p$. [374, 394] In the framework of this theory, the fraction of the remanent polarization of a porous material compared to a dense counterpart is expressed as:

$$P_r/P_r^0 = d_p(p, E_{\text{app}}) \cdot (1-p) \qquad \text{Equation 4}$$

The local electric field in some volumes of the material is too low to initiate domain wall movement (Figure 14a). Consequently, the depolarization factor and thus the volume fraction of active material can be tuned by tailoring the local electric field distribution, e.g., through microstructural engineering the degree of porosity or its shape [101, 374] or by increasing the applied electric field [101]. A higher applied electric field will shift the local electric field distribution to higher values and will activate previously inactive volumes. Adjusting the externally applied electric field thus offers the potential to tailor the degree of poling and the piezoelectric response of porous ferroelectrics/ferroelastics.

Another consequence of the gap between experimentally measured and theoretically predicted values of the remanent polarization in Figure 15d is, that a direct comparison between the coercive fields obtained from polarization measurements of different porous materials should be undertaken with care, since the polarization reversal remains incomplete in most cases (Figure 15d). This may explain the previously revealed discrepancies in trends of the coercive fields of porous materials with increasing degree of isometric porosity [374]. Some authors claim that the coercive field decreases with increasing total porosity, [372, 387, 395] others find an increase, [389, 394] while again others observe no effect [386, 390]. Other reports find that the coercive field first decreases before it increases again. [374, 396] Calculations predict a more realistic increase of the coercive field with increasing total porosity if a sufficiently high electric field is applied to activate the entire material. [397, 398]

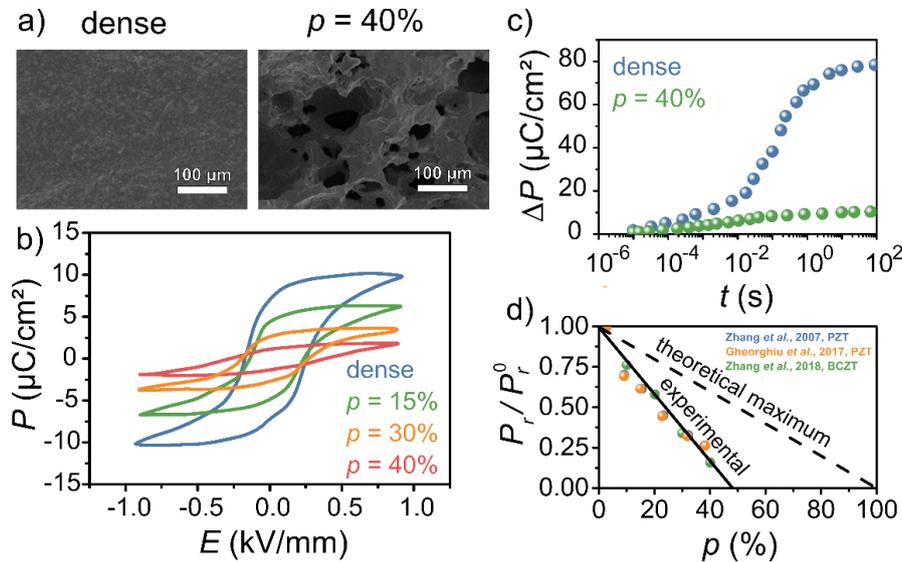

*Figure 15: Impact of the degree of isometric porosity in domain wall dynamics. a) Micrographs of a dense material and a material with isometric pores (p=40%) of a BCZT material are displayed. The impact of porosity on the polarization hysteresis loops of porous BCZT materials with different degrees of porosity is displayed in b). Reprinted from [374], with permission from Elsevier. c) Time-dependent development of the switched polarization, ΔP, of a dense and a porous (p=40%) PZT ceramic material. [128] d) Dependence of the remanent polarization, $P_r$, of selected porous materials with porosity, p, normalized to the remanent polarization of a dense counterpart, $P_r^0$. [372, 374, 387] The experimental data of the selected porous materials follow the solid line. The theoretical maximum, which displays the real available material, is displayed as the dashed line.*

4.3.2. Orientation angle of anisometric pores

Materials with oriented anisometric pores can be obtained by freeze casting [399, 400] or paper-derived approaches [393, 401]. The impact of pore orientation angle with respect to the electric field on domain wall dynamics and performance has been systematically studied in ref. [101]. Micrographs of a freeze-cast polycrystalline PZT ceramic are displayed in Figure 16a. The individual porous channels are magnified in Figure 16b, highlighting the high degree of pore alignment achievable by freeze-



casing. Macroscopic polarization and strain loops of samples with different orientation angles of anisometric pores, $\theta$, with respect to the electric field are displayed in Figure 16c and d. A dense counterpart of the same grain size is displayed for comparison. Increasing $\theta$ away from the direction of the applied electric field results in a reduction of the remanent and saturation polarization, which comes along with a reduction in bipolar strain. This effect was related to statistical local electric field distributions, (Figure 14c) shifting towards lower values and broadening with increasing $\theta$ away from the applied electric field. [101, 128, 371, 373, 376] This reduces the amount of actively switched material, and thus the obtainable remanent polarization and bipolar strain. Interestingly, at the same time, the sample with anisometric pores oriented parallel to the applied electric field ($\theta = 0°$) has comparable electromechanical properties to a dense counterpart. This is a consequence of the comparable local electric field distributions, which allows to activate the porous material with pores oriented parallel to the applied electric field nearly completely (Figure 14c). In this frame small signal piezoelectric coefficient and permittivity decrease with increasing pore orientation angle, which is the key to obtain high figures of merit for sensing ($g_{33}$) and energy harvesting ($d_{33} \cdot g_{33}$), as summarized in Figure 16d. [371, 399, 402]

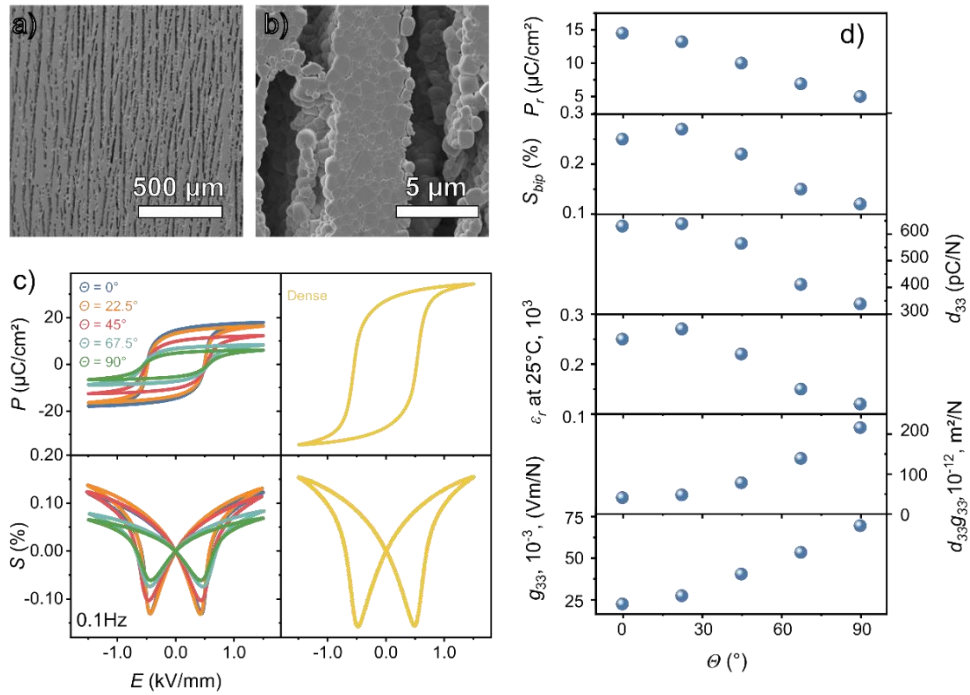

*Figure 16: Impact of pore orientation angle on domain wall dynamics and functional properties. The microstructure of porous (34 vol.%) PZT samples containing anisometric pores is displayed in a) and a magnification of the porous channels is provided in b). Macroscopic polarization and strain loops are displayed in c) for porous materials with different pore orientation angles. The angle $\theta$ represents the pore orientation angle with respect to the external electric field. The response of a dense polycrystalline material with the same grain size is displayed for comparison. The dependence of functional properties (remanent polarization, $P_r$, bipolar strain, $S_{bip}$, small-signal piezoelectric coefficient, $d_{33}$, relative permittivity, $\varepsilon_r$, figure of merit for sensing, $g_{33}$, figure of merit for energy harvesting, $d_{33}g_{33}$) on the pore orientation angle $\theta$ is displayed in d).*

### 4.3.3. Summary

Experiments and simulations have revealed the role of porosity on the local electric driving forces for domain wall movement. While the impact of the degree of total porosity and the pore orientation angle on domain wall dynamics and functional properties have been heavily researched, experimental studies on the impact of pore radius are lacking. To further reveal the impact of mechanical driving forces on the static domain structure and domain wall dynamics, imaging techniques on porous materials are highly desirable. Orienting highly anisotropic pores has been further identified as an effective mean to engineer domain wall dynamics via local electric field distributions. Beyond the microstructural approaches discussed in section 4.3, combining porous ferroelectric/ferroelastic materials with other microstructural engineering parameters, such as grain size (section 4.2.1) and crystallographic texturing (section 4.2.2) further broaden the playground for property engineering. Freeze casting in combination with the templated grain growth process, for example, provides full control over the pore orientation together with the orientation of the grains. [403] While first materials were manufactured, the dielectric and piezoelectric performance has not yet been characterized. Processing routines to control the porosity together with the grain size are also on the way, [404] given porous ferroelectric/ferroelastic materials a new dimension to engineer the functional performance via microstructural engineering.

## 5. Computational approaches

In this section we summarize selected computational approaches, which allow insight on different aspects of the relationship between domain wall dynamics and structure/microstructure in ferroelectric/ferroelastic polycrystals. We start in section 5.1 with phase field simulations, which predict the relationship on the mesoscale. Complementary, stochastic (section 5.2) and micromechanical modelling (section 5.3) give insights into time and field-dependent response of ferroelectric/ferroelastic materials on the macroscopic scale. Going beyond the approaches summarized in this review, the domain wall dynamics utilizing effective Hamiltonian approaches and molecular dynamic simulations are discussed in refs. [405-407], while a more general overview of atomistic models can be found in refs. [408, 409]. Homogenized energy models [74, 410-412] or atomistic dummy models [413, 414] allow to study further specific aspect of domain wall dynamics in polycrystalline ferroelectrics/ferroelastics.

### 5.1. Phase field method

The phase-field method has been established as an effective tool to predict the domain structure evolution and its correlations with material properties and behaviors in ferroelectric ceramics, single crystals, and thin films. [415, 416] The method is based on the thermodynamics and kinetics of materials at the mesoscale. The total free energy of a ferroelectric polycrystal includes contributions from the bulk, elastic, electric, and domain wall energies. [415, 416] To describe more specific aspects of



ferroelectrics, it is also possible to include other contributions, e.g., from charged defects [417], flexoelectric coupling [418], and antiferrodistortive ordering [419]. Table 2 summarizes these energy expressions and the material parameters together with possible sources of experimental methods and/or density functional theory (DFT) calculations to obtain material parameters needed in phase-field simulations. In the phase-field model of polycrystalline ferroelectric/ferroelastic materials, all tensor parameters are grain orientation-dependent and thus the properties within different grains are related by the corresponding matrix rotation operation. [420] However, the structures and properties of grain boundaries are largely unknown and thus require future extensive experimental and theoretical exploration. [316] Phase-field simulations can be employed to model and predict the domain structure evolution, as well as the equilibrium domain structures under externally applied thermal, mechanical, and electric stimuli by numerically solving the time-dependent evolution equations for polarization along with mechanical and electrostatic equilibrium equations [415, 416].

Phase-field modelling has been extensively utilized to model the equilibrium domain structures [421, 422], domain dynamics [423, 424], mechanical behavior [425], piezoelectric responses [426, 427], and electrostatic energy-storage properties [367]. There have also been phase-field simulations to understand the effect of grain morphology [428], crack propagation [429], and mobile defects [417, 430-433] such as oxygen vacancies in ferroelectric ceramics. The models found that grain boundaries, structural, and charge defects in ferroelectric polycrystals result in strongly inhomogeneous internal electric and stress fields, leading to complex polar configurations and material behavior considerably different from their single-crystalline counterparts. [434, 435]

Exemplary modelling results for the impact of grain size in a polycrystalline BT materials and degree of crystallographic texture in a polycrystalline PMN-PT are displayed in Figure 17a and b, respectively, and will be discussed with respect to the experimental results presented in sections 4.2.1 and 4.2.2. In agreement with experimentally measured polarization loops displayed in Figure 8a, phase field simulations predict slimmer polarization loops for ceramics with larger (100 nm) grain sizes than that smaller (20 nm) ones. Furthermore, the grain size dependence of the coercive field and the remanent polarization obtained from phase-field simulations are in good qualitative agreement with experimental results discussed in Figure 9a. However, phase field simulations do not yet cover all experimentally observed aspects, e.g., the peak in the small signal permittivity and piezoelectric coefficient observed for BT in Figure 9c. The dependence of the piezoelectric coefficient on the degree of crystallographic texture in PMN-PT is displayed in Figure 17b. [428] The experimental piezoelectric response of a PMN-PT material is presented for comparison. [436] The results obtained from phase field simulations largely agree with experimental observations, as discussed in section 4.2.2.2. Figure 17b further displays the spatial distribution of local electric field and mechanical stress concentrations obtained from simulations, highlighting concentrations at the boundary between misoriented grains.

The advantage of phase-field modeling in understanding the experimental observations at the mesoscale lies in its ability to resolve multi-physical fields and quantify different energy contributions. More details about the phase-field model of ferroelectrics and its applications, challenges, and prospects are reviewed in ref. [415]. Phase-field method can be employed to not only help to understand experimental observations and measurements, but also to guide the design of materials, e.g., ultrahigh piezoelectric response in relaxor ceramics [437], enhanced dielectric response in tricritical ferroelectrics [438], enhanced energy-storage performance in solid-solution polycrystalline films [439], and giant electrocaloric effect in relaxor ceramics-polymer nanocomposites [440]. Moreover, phase field simulations have been employed to reveal the impact of dislocations, [253] precipitates [441] or more complex topological defects [176] on the domain structures in ferroelectric materials.

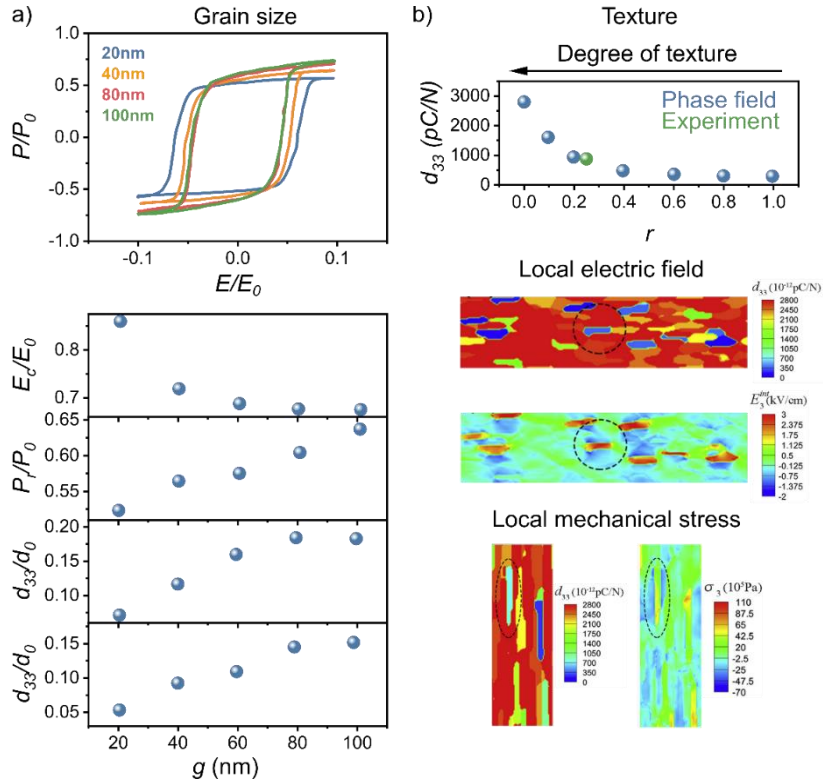

*Figure 17: Insights obtained by phase field modelling on the impact of microstructure on domain wall dynamics and functional properties of polycrystalline ferroelectric/ferroelastic materials: a) Simulated impact of grain size on polarization hysteresis in polycrystalline BT. [423]. b) Impact of degree of crystallographic texture (quantified by the March Dollas parameter, r [442]) on piezoelectric properties in PMN-PT polycrystalline ceramics. The experimental reference value, obtained from ref. [436], is plotted for comparison. To highlight local electric and mechanical stress concentrations, misoriented grains are indicated by circles in b). Reprinted from [428], with permission from Elsevier.*



**Table 2:** *Energy terms and necessary material parameters in the phase-field model of ferroelectrics. Possible approaches to obtaining the material parameters are also provided.*

| Energy Term | Expression | Variables | Key Parameters and Possible Obtainments |
|---|---|---|---|
| $f_{\text{bulk}}$ | $\alpha_i P_i^2 + \alpha_{ij} P_i^2 P_j^2 + \alpha_{ijk} P_i^2 P_j^2 P_k^2$ | $P_i$: Polarization; $\alpha_i, \alpha_{ij}, \alpha_{ijk}$: Landau Coefficients; | $\alpha_i, \alpha_{ij}, \alpha_{ijk}$ relating to different-order dielectric stiffness [443, 444] |
| $f_{\text{elastic}}$ | $\frac{1}{2} c_{ijkl}(\varepsilon_{ij} - Q_{ijmn}P_m P_n)(\varepsilon_{kl} - Q_{klmn}P_m P_n)$ | $\varepsilon_{ij}$: total strain; $c_{ijkl}$: stiffness tensor coefficients; $Q_{ijkl}$: electrostrictive coefficients; | $c_{ijkl}$, $Q_{ijkl}$ inelastic neutron scattering [445]; plate resonance measurements [446]; DFT calculations [447] |
| $f_{\text{electric}}$ | $-P_i E_i - \frac{1}{2}\varepsilon_0 \kappa_{ij}^b E_i E_j$ | $E_i$: electric field; $\kappa_{ij}^b$: background permittivity | $\kappa_{ij}^b$, high-temperature dielectric measurement [448] |
| $f_{\text{gradient}}$ | $\frac{1}{2}\gamma_{ijkl}\frac{\partial P_i}{\partial x_j}\frac{\partial P_k}{\partial x_l}$ | $\gamma_{ijkl}$: gradient energy coefficients | $\gamma_{ijkl}$, neutron diffuse scattering experiment [446] |
| $f_{\text{defects}}$ | $nE_C - pE_V + N_i E_i - T(s_n + s_p + s_{trap})$ | $E_C$: conduction band bottom; $E_V$: valence band top; $n$: electron concentration; $p$: hole concentration; $N_i, E_i$: concentration and electron level of the $i$th defect species; $s$: entropy | $E_C, E_V, N_C, N_V, E_i$ electrical conductance measurements [449-451] |
| $f_{\text{flexo}}$ | $\frac{1}{2}F_{ijkl}\left(\frac{\partial P_k}{\partial x_l}\sigma_{ij} - \frac{\partial \sigma_{ij}}{\partial x_l}P_k\right)$ | $F_{ijkl}$: flexoelectric coefficients; $\sigma_{ij}$: stress | $F_{ijkl}$: neutron diffuse scattering [452, 453] |
| $f_{\text{antiferrodistortion}}$ | $\beta_i q_i^2 + \beta_{ij} q_i^2 q_j^2 + \beta_{ijk} q_i^2 q_j^2 q_k^2$ | $q_i$: structural order parameter; $\beta_i, \beta_{ij}, \beta_{ijk}$: Landau Coefficients; | $\beta_i, \beta_{ij}, \beta_{ijk}$; temperature-dependent Raman spectrum measurement [454] |

### 5.2. Stochastic modelling

Stochastic models are based on the theories developed by Johnson and Mehl [455] and independently by Avrami [456]. Isibashi and Takagi systematically applied the theoretical approaches to study domain wall dynamics in ferroelectric single crystals, [115] using the mathematical concept of Kolmogorov [457], which is commonly used to describe crystallization kinetics in metals. According to the classical Avrami approach, the driving force for domain wall movement is the difference in the free energies between the starting and the new phases. In the framework of the Kolmogorov-Avrami-Ishibashi (KAI) model, it is assumed that the mean size of individually transformed regions is small in comparison to the sample size, that the nucleation probability for new domains is spatially uniform, and that the value of domain wall velocity is the same in all regions. Additionally, the domain wall movement is considered as the time-dependent step. [114] In the KAI model, a temporal dependence of the reversed polarization is presented as

$$\Delta P(t, \tau_{E_{Sw}}) = \Delta P_{max}\left\{1 - \exp\left[-\left(\frac{t}{\tau_{E_{Sw}}}\right)^\zeta\right]\right\}. \qquad \text{Equation 5}$$

Here, $\tau_{E_{Sw}}$ is a unique time constant related to the movement of the domain wall after nucleation, $\Delta P_{max}$ is the amount of the reversed polarization reached at saturation, and $\zeta$ is the Avrami exponent. The latter depends on the dimensionality of the domain and is expected to take only integer values from 1 to 4. In this frame, it is distinguished between one-step-nucleation (no additional nuclei form and the domain walls move with a constant velocity, $\zeta = \delta_{\text{dim}}$) and continuous nucleation (additional nuclei form during the switching process with a constant rate and domain walls move with a constant velocity, $\zeta = \delta_{\text{dim}} + 1$). The value $\zeta$ can be related to the dimensionality of the growing domain $\delta_{\text{dim}}$. Stripe domains ($\delta_{\text{dim}} = 1$), circular domains ($\delta_{\text{dim}} = 2$), and spherical domains ($\delta_{\text{dim}} = 3$) are distinguished (Figure 18a). The correlation between the exponent $\zeta$ and the dimensionality of the growing domain has not yet been confirmed experimentally.

    The KAI model is commonly used to describe the time-dependent polarization evolution of single crystalline ferroelectrics/ferroelastics. Studied materials include BT [126, 362], $(CH_3NH_3)_5Bi_2Br_{11}$ [458, 459], TGS [460, 461], and $KTiOPO_4$ (KTP) [462]. A representative example of the time-dependent switched polarization of a [111] oriented BT single crystal is provided in Figure 18b. Here, the solid lines reveal fits according to Equation 5. The Avrami exponents were found to be field-dependent and non-integer. [362] To account for non-integer Avrami coefficients, the KAI model was modified to take finite sample sizes [463, 464] into account and a new mathematical treatment was suggested. When the growing domain touches a boundary, it stops growing in that direction, resulting in a time-dependent change of the Avrami exponent, referred to as "geometrical catastrophe" [39, 465]. Non-integer Avrami coefficients were also explained by a mixing of one step and continuous nucleation [462].

    The KAI model was applied to describe domain wall dynamics in more complex systems, characterized by a higher degree of disorder than single crystals. For example, domain wall dynamics in certain copolymers revealed values as large as $\zeta = 5$. [466] Equation 5 was utilized to describe the domain wall dynamics also in polycrystalline [467-469] and epitaxial [470, 471]



thin films. By comparing the switching dynamics of polycrystalline thin films to single crystalline counterparts, it was found that thin films exhibit a broadened distribution of switching times. [472] This finding led to an extension of the KAI model by an aggregation of regions with different switching dynamics, while the domain wall dynamics in the entire sample is a superposition of all these regions. [473] In the nucleation limited switching (NLS) model, switching of one region is limited by the nucleation of new domains rather than by domain wall movement, and a smooth and exponentially broad distribution of nucleation waiting times was suggested to describe the switching process in a polycrystalline thin film best. Individual regions thereby correspond to single grains or clusters of grains and grain boundaries act as frontiers to limit the propagation of the switched region. [322, 474] This extended model was successfully applied to study polarization reversal dynamics in polycrystalline thin films. [475, 476]

Building up on the NLS model, domain wall dynamics in polycrystalline bulk ceramics were described by the inhomogeneous field mechanism (IFM) model. The IFM model assumes that the distribution of switching times is related to an inhomogeneous distribution of the local electric field, as discussed in section 2.2.2. [83, 117] In this context, the time-dependent switched polarization

$$\Delta P(E_{Sw}, t) = \Delta P_{max} \int_0^{E_{Sw}/E_{max}(t)} \frac{du}{u} \Phi(u) \qquad \text{Equation 6}$$

can be calculated for any high voltage field pulse $E_{Sw}$, if the functions $E_{max}(t)$ and $\Phi(u)$, which represent a fingerprint of the domain wall dynamics of the ferroelectric material, are known. The functions $E_{max}(t)$ and $\Phi(u)$ can be extracted from polarization dynamic measurements performed at different field pulses, $E_{Sw}$, as described in refs. [83, 124]. This model was applied to study the impact of crystallographic structure [123-125], degree of crystallographic texture, [341] and porosity [128] on domain wall dynamics in polycrystalline ferroelectrics/ferroelastics and is even applicable to ferroelectric polymers. [477, 478]

Experimental advances allowed to measure the dynamics of macroscopic strain simultaneously with the switched polarization, which allows to distinguish 180° and non-180° domain wall dynamics (section 2.2 and Figure 2). [89, 100] These findings helped to improve stochastic models further. In contrast to the classical KAI approach, the multi-step stochastic mechanism (MSM) [91] model includes two parallel channels of switching for materials with a tetragonal symmetry: 180°-polarization reversal and sequential two-step 90° switching events. This allows to calculate the time-dependence of the switched polarization and strain, as schematically indicated in Figure 18d. Here, experimental curves are highlighted by red symbols, while calculated contributions from 180° and 90° switching events are displayed by dotted lines. The sum is represented by the black solid line and describes the experimental observation well. An important finding of the MSM model is, that 90° switching events account for 34% of the switched polarization in a tetragonal $Pb_{0.985}V_{Pb_{0.005}}La_{0.01}(Zr_{0.475}Ti_{0.525})O_3$ polycrystalline ceramic. In the framework of the MSM model, the domain dynamics in polycrystalline ferroelectrics/ferroelastics are related to the dimensionality of the growing domain [91], as well as the inhomogeneous distribution of the local electric field [118]. An extension of the MSM model to polycrystalline orthorhombic [479] and rhombohedral [235] materials also exists.

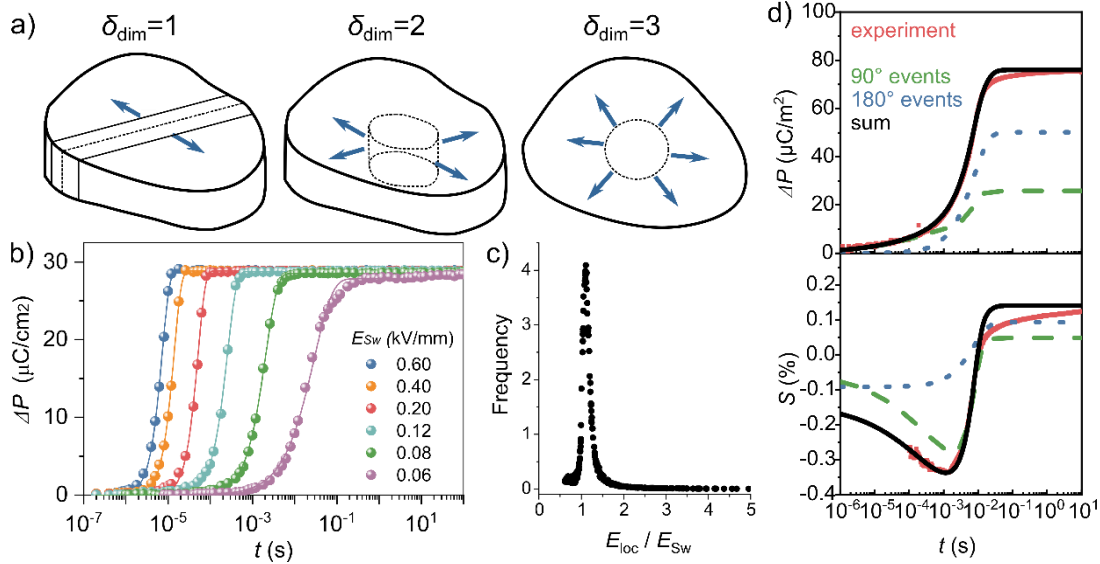

*Figure 18:* The description of domain wall dynamics in polycrystalline ferroelectrics/ferroelastics using stochastic modelling. a) The shape of domains with different dimensionality according to the KAI approach are visualized. [115] The blue arrows indicate the direction of growth. b) Switched polarization as a function of time for a [111] oriented BT single crystal. The symbols correspond to the experimental results obtained at a constant switching field of pulse height $E_{Sw}$, as indicated, while the solid lines represent fits according to Equation 5. [362] c) Weighted statistical distribution of the local electric field of a polycrystalline PZT ceramic sample obtained from the IFM Model (Equation 6) [83]. d) Switched polarization and strain variation as a function of time for a tetragonal $Pb_{0.985}V_{Pb_{0.005}}La_{0.01}(Zr_{0.475}Ti_{0.525})O_3$ polycrystalline ceramic at a constant applied electric field. Fits to the experimental data according to the MSM model display the calculated contribution of 90° and 180° switching events using the MSM model. [91, 100]

5.3. Micromechanical modelling

In early micromechanical models, [480, 481] the polycrystalline material was constructed by many randomly oriented tetragonal and single-domain grains and the switching behavior of each individual grain was described by an idealized rectangular hysteresis behavior. To simulate the response of the polycrystalline material, each grain was subjected to the same externally applied field and stress. The applied field was increased stepwise and for each step the change in energy, related to a possible 180° or 90° switching event, was calculated. When the sum of the electric and the mechanical work exceeded a critical value, the direction of the spontaneous polarization has switched:

$$E_i \Delta P_i + \sigma_{jk} \Delta S_{jk} \geq 2 P_s E_c \qquad \text{Equation 7}$$

Here, $\Delta P_i$ is the change in remanent polarization and $\Delta S_{jk}$ is the change in remanent strain of the switching domain. The first set of micromechanical models assumed a very simplified behavior. Individual grains were described as monodomain states and domain walls were neglected. [481-483] Also, inhomogeneities in the electric field and mechanical stress, which are both a key feature of polycrystalline materials (section 2.2), were ignored. [484] Later, interactions between grains were implemented via the Eshelby approach [303, 485, 486] or finite element solutions. [487] The models were further extended and domain configurations [488, 489] and interaction between domains [490] were taken into account. Beyond tetragonal materials, extensions to rhombohedral crystal structures [486] or materials with a phase coexistence [491] are available. For a more detailed description of micromechanical models, we refer the reader to ref. [492].

## 6. Conclusions, current challenges, and future perspectives

Domain wall movement has been identified as one of the main contributions to the dielectric and electromechanical properties of ferroelectric/ferroelastic materials. While the dynamics of the walls can be tuned via the defect chemistry, the impact of structure and microstructure on the functional properties of polycrystalline ferroelectric/ferroelastic materials has been intensively studied. In this frame, polycrystals offer many degrees of freedom, including crystallographic structure, degree of crystallographic orientation, grain size, and porosity. The impact of each of these parameters on ferroelectric/ferroelastic domain wall movement needs to be understood. The main challenge is to disentangle the high interdependency between parameters governing the formation of the static domain structure and domain wall dynamics. In this review article, we address this challenge by explaining domain formation and domain wall movement by electric and mechanical driving forces, which have been revealed to govern domain wall dynamics in polycrystalline ferroelectrics/ferroelastics. [89-91] To generate a comprehensive understanding of the complex structure-property relationship in ferroelectric/ferroelastic materials in the future, dielectric and electromechanical measurements need to be combined with advanced structural and nanoscale dynamic characterization.

### 6.1. Advanced characterization methods

High-energy diffraction and advanced SPM techniques are particularly interesting for this purpose. Advanced X-Ray microscopy techniques visualize and quantify domain wall dynamics, spatially resolved on the nanoscale. State-of-the-art synchrotron sources achieve spatial resolutions down to sub-100 nm [171] and time-resolutions in the nanosecond-range in stroboscopic modes were reported in bulk geometry. [493] Furthermore, X-Ray free electron lasers (XFEL) enable ultrafast time resolutions, which enable capturing polarization rotation phenomena occurring at picoseconds. [494] In this frame, it is also known that ferroelectric materials are suffering from radiation damage. [495-497] Defined boundary conditions, such as the interplay between radiation time and electric field amplitude or the influence of different chemical elements on the interactions between ferroelectric/ferroelastic materials and high-energy synchrotron radiation therefore need to be established.

Besides X-Ray diffraction, PFM-based techniques and spectroscopic approaches are highly attractive to get insights into the ferroelectric/ferroelastic domain structure, since they offer nanoscale spatial resolution. The ferroelectric/ferroelastic domain size and domain morphology can be quantified together with the local coercive voltages, relaxation times or Rayleigh behavior. For polycrystalline materials, which feature a complex domain structure and a random orientation of the individual crystallites, advances towards automated SPM experiments and the use of machine learning approaches are of particular interest. [498] Such measurements provide the possibility to quantify the dynamics of individual types of domain walls, providing a new pathway to study the impact of microstructure on domain wall dynamics in polycrystalline ferroelectrics/ferroelastics. [499]

The interplay between microstructural features and chemical defects, e.g., space charge layers in the vicinity of grain boundaries, [500] and their impact on domain structure, are largely unexplored. Atom probe tomography (APT) is a common technique applied in metallurgy to get insights into objects such as precipitates or dislocations. [501] It combines a high chemical sensitivity and accuracy with three-dimensional spatial resolution, facilitating to map and quantify subtle changes in defect concentrations at grain boundaries in ferroelectric materials, which is the key to reveal the impact of defects of atomic size on the local domain structure and domain wall dynamics.

### 6.2. New microstructural approaches using extended defects

Going beyond the structural and microstructural tools introduced in Figure 4a and b, new approaches are well on the way now. The local interaction of domain walls with strain fields originating from designed extended defects, such as secondary phases, [502, 503] dislocations, [253, 504] or precipitates [441] have been outlined, providing new levers to control the static domain structure, domain wall back-switching or domain wall pinning in polycrystalline ferroelectrics/ferroelastics. Pioneering studies on BT single crystals, for example, outline a giant electric-field dependent permittivity (5800) and large-signal piezoelectric coefficient (1890 pm/V), indicating that controlling the domain wall dynamics via mechanical dislocation imprint is a powerful tool for functional property engineering. [253]

Besides strain fields, advances in nanoscale electric characterization have created a fundamental understanding on the electronic transport properties of ferroelectric domain walls. [153, 505] Bulk ferroelectric materials will strongly benefit from the ongoing advanced in characterization and understanding of the electronic response on the domain wall level. First intriguing examples suggest domain wall networks as reconfigurable barrier layer capacitors in single crystalline ErMnO$_3$. [506, 507] In addition, the unusual low frequency enhanced electromechanical properties in polycrystalline BiFeO$_3$ were explained by conductive domain walls [508, 509]. New physical phenomena, e.g., the uncoupling of lattice strain and domain wall switching [510], not available in classical materials such as PZT or BT, in addition provide new degrees of freedom to control the frequency-dependent properties of bulk ferroelectric materials. While charged ferroelectric domain walls have been identified as a general phenomenon appearing in many ferroelectrics, [511] utilizing their collective electronic transport behavior in bulk application in polycrystals is unchartered territory. An outlook on the application potential of charged ferroelectric domain walls in macroscopic capacitor or actuator applications is provided in ref. [153].

### 6.3. Ferroelectric materials with more complex correlation phenomena

Going beyond classical ferroelectricity towards more complex degrees of ferroelectric order, improper ferroelectric materials have attracted attention. [512-514] In contrast to the proper systems discussed so far, the spontaneous polarization is not the primary order parameter in improper ferroelectrics. [515, 516] Ternary hexagonal manganites, $R$MnO$_3$ ($R$ = Sc, Y, In, and Dy – Lu) are one example of improper ferroelectrics. In $R$MnO3, the polarization originates as a symmetry enforced by-product of a structurally driven phase transition. [517] As a consequence, the materials exhibit a large variety of unusual physical phenomena at the level of ferroelectric domains. Intriguing examples range from domain walls with unique functional electronic properties [512-514, 518, 519] to topological vortex structures that have been utilized to test cosmological scaling laws [520-522]. Most of the research so far, however, focused on single crystals and thin films, whereas only few studies were performed on polycrystalline systems. [523-526] In particular the possibility to utilize the microstructure to engineer the ferroelectric domain structure via naturally occurring topologically protected vortex/anti-vortex pairs remains to be explored. A novel domain scaling behavior in polycrystalline ErMnO$_3$ materials was discovered, demonstrating that the established scaling behavior in BT or PZT (Figure 9e and f) can be suppressed and inverted. [176] This behavior was related to the interaction of topologically protected vortex/anti-vortex pairs with long-ranging elastic strain fields. Combining microstructural engineering with improper ferroelectrics provides an exciting playground to engineer the functional behavior of polycrystalline ferroelectrics via the domain structure and domain wall dynamics in the future.

**Acknowledgements**

J.S. acknowledges the support of the Feodor Lynen Research Fellowship of the Alexander von Humboldt Foundation and NTNU Nano for the support through the NTNU Nano Impact fund. This work was partially supported by the Deutsche Forschungsgemeinschaft (DFG) under the Grant Nr. 270195408 (KO5100/1-1). L.Q.C. would like to acknowledge the Alexander von Humboldt Foundation for the Humboldt Research Award and the US National Science Foundation under the grant number DMR-1744213.